\documentclass[11pt,letterpaper,nofootinbib]{revtex4-2}
\usepackage[top=100pt,bottom=0pt,letterpaper]{geometry}
\usepackage{adjustbox}
\usepackage[T1]{fontenc}
\usepackage{lmodern}
\usepackage[utf8]{inputenc}
\usepackage{microtype}
\usepackage{setspace}
\usepackage{float}
\restylefloat{table}

\usepackage{dcolumn}
\usepackage{lipsum}
\usepackage{graphicx,epsfig}
\usepackage{psfrag}
\usepackage{bm}
\usepackage{epstopdf}
\usepackage{latexsym}
\usepackage{multirow}
\usepackage{amsmath,amssymb}
\usepackage{enumerate}
\usepackage{hyperref}
\usepackage[FIGTOPCAP]{subfigure}
\usepackage{empheq,mathtools}

\usepackage{booktabs}
\usepackage{tabularx}
\usepackage{longtable}
\def\be {\begin{equation}}
\def\ee {\end{equation}}
\def\ba {\begin{eqnarray}}
\def\ea {\end{eqnarray}}

\def\bc {\begin{center}}
\def\ec {\end{center}}
\newcommand{\bdm}{\begin{displaymath}}
\newcommand{\edm}{\end{displaymath}}

\def\l  {\lambda}

\def\O  {\Omega}

\def\th {\theta}
\def\Th {\Theta}

\def\S {\Sigma}

\def\ra {\rightarrow}

\def\bi {\begin{itemize}}
\def\ei {\end{itemize}}

\def\> {\rangle}
\def\< {\langle}

\def\bc {\begin{center}}
\def\ec {\end{center}}

\usepackage{amsfonts}

\begin{document}
\title{Geodesic Congruences in 5D Warped Ellis-Bronnikov Spacetimes}

\author{Vivek Sharma} \email[email: ]{svivek829@gmail.com}
\affiliation{Department of Physics, Indira Gandhi National Tribal University, Amarkantak - 484887, India}

\author{Suman Ghosh} \email[email: ]{suman.ghosh@bitmesra.ac.in}
\affiliation{Department of Physics, Birla Institute of Technology, Ranchi - 835215, India}

\begin{abstract}

We study the timelike geodesic congruences in the generalized Ellis-Bronnikov spacetime (4D-GEB) and in  recently proposed 5D model where a 4D-GEB is embedded in a warped geometry (5D-WGEB) and conduct a comparative study. Analytical expressions of ESR variables (for 4D geometries) are found which reveal the role of the wormhole parameter. In more general 4D and 5D scenarios geodesic equation, geodesic deviation equation and Raychaudhury equations are solved numerically. The evolution of cross-sectional area of the congruences of timelike geodesics (orthogonal to the geodesic flow lines) projected on 2D-surfaces yield an interesting perspective  and shows the effects of the wormhole parameter and growing/decaying warp factors. Presence of warping factor triggers rotation or accretion even in the absence of initial congruence rotation. Presence of rotation in the congruence is also found to be playing a crucial role which we discuss in detail.

\end{abstract}


\maketitle
\section{introduction} \label{sec:1}

Einstein and Rosen introduced the concept of wormholes, as early as in 1935 \cite{Einstein:1935tc}, which generate `short-cuts' in spacetime and allow `apparently faster than light' travel \citep{Visser:1995cc, Hawking:1973uf, Wald:1984rg, lobo2017wormholes, Witten:2019qhl} between two distinct spacetime points, dubbed as the Einstein-Rosen bridge. Wheeler, later, referred these solutions as `wormholes' \cite{Wheeler:1957mu}. It was soon realised that the wormhole solutions do not form a stable structure-- it's `throat' closes up too quickly when subjected to even tiny perturbation \citep{Kruskal:1959vx, Fuller:1962zza, Eardley:1974zz, Wald:1980nm}. Morris and Thorne showed that one can get a `traversable' wormhole solution by threading exotic form of matter (matter with negative energy density that violates null energy condition (NEC)) at or near the throat \citep{Morris:1988cz, Visser:1995cc}. Remarkably, the concept of exotic matter (such as dark energy, dark matter and phantom energy) is found to be useful in explaining several cosmological observations such as: the late time accelerated expansion of the universe, behaviour of the galactic rotation curves and the mass discrepancy in clusters of galaxies \cite{Lobo:2008sg}. It seems that the natural habitat for such matter would be of quantum origin. However, it has been noted that approaches using quantum features of standard model matter are insufficient for creating macroscopic wormholes \cite{Witten:2019qhl}.

Despite the facts stated earlier, certain classical ways are devised to circumvent the necessity for matter with negative energy density that violates energy conditions \cite{Hochberg:1990is, Bhawal:1992sz, Agnese:1995kd, Samanta:2018hbw, Fukutaka:1989zb, Ghoroku:1992tz, Furey:2004rq, Bronnikov:2009az}. Alternative theories of gravity, often known as the modified gravity theories, also offer new ways to avoid energy condition violations. Even though the convergence condition of null geodesics is violated in some cases, the literature has a considerable number of non-exotic matter models under modified gravity \citep{Lobo:2008zu, Kanti:2011jz, Kanti:2011yv, Zubair:2017oir, Shaikh:2016dpl, Ovgun:2018xys, Canate:2019spb}. So far, the $f(R)$, $f(R, T)$, $f(Q)$, and higher order gravity theories have garnered sufficient attention. It has been demonstrated that cosmic acceleration can be explained using $f(r)$ gravity \cite{Nojiri:2003ft} and that these models can provide wormhole solutions with viable matter sources \citep{Lobo:2009ip, Garcia:2010xb, MontelongoGarcia:2010xd, Sajadi:2011oei, Moraes:2017dbs, Sahoo:2017ual, Moraes:2019pao, Sahoo:2020sva, Hassan:2021egb, Mustafa:2021ykn}.

Although wormholes are still regarded speculative, new advancements in precision black hole measurements have enhanced the need of evaluating plausible wormhole models (as black hole mimickers). Studies of events such as wormhole merger \citep{Krishnendu:2017shb, Cardoso:2016oxy} and their quasinormal modes \citep{Aneesh:2018hlp, DuttaRoy:2019hij} can aid in the detection of wormhole signature in the cosmos. Similarly. their lensing effects, shadows, Einstein ring and other features may be analysed in detail for possible detection \citep{Abe:2010ap, Toki:2011zu, Takahashi:2013jqa, Cramer:1994qj, Perlick:2003vg, Tsukamoto:2012xs, Bambi:2013nla, Nedkova:2013msa, Zhou:2016koy,  Dzhunushaliev:2012ke, Dzhunushaliev:2013lna,  Aringazin:2014rva, Dzhunushaliev:2016ylj}. Any of these signatures, if detected, may also favour modified gravity theories over general relativity.

The Ellis-Bronnikov wormhole (4D-EB) \citep{Ellis:1973yv, Bronnikov:1973fh} was constructed separately by Ellis and Bronnikov employing phantom scalar field (a field with negative kinetic term) and is one of the most researched wormhole geometries. Several studies on this spacetime can be found in the literature, including geometry of spinning 4D-EB spacetime \cite{Chew:2016epf}, generalized spinning of 4D-EB wormhole in scalar-tensor theory \cite{Chew:2018vjp}, hairy Ellis wormholes solutions \cite{Chew:2020svi}, Ellis wormholes in anti-de Sitter space \cite{Blazquez-Salcedo:2020nsa}, stability analysis of 4D-EB solution in higher dimensional spacetime \cite{Torii:2013xba} as such. Kar et al. presented a generalised version of 4D-EB spacetime (4D-GEB) \cite{Kar:1995jz}, where the necessity for exotic matter is {\em partially} evaded by introducing a new wormhole parameter, $m \geq 2$ ($m = 2$ corresponds to 4D-EB geometry). Quasinormal modes, echoes and some other aspects of 4D-GEB spacetimes is reported in \cite{DuttaRoy:2019hij}. We recently suggested a further generalisation \cite{Sharma:2021kqb} where the 4D-GEB geometry is embedded in a five-dimensional warped spacetime and showed that energy conditions are satisfied even for $m=2$ i.e. a novel 5D-EB geometry.
Such embedding is also considered in the context of Schwarzschild-like spacetime \cite{Culetu:2021zun} as well. Recently Kar has proposed another 5D warped wormhole model where the warping chosen is largely inspired by the non-static Witten bubble \cite{Kar:2022omn}.

The theories of extra spatial dimensions started with the work of Kaluza (1921) and Klein (1926) \cite{Kaluza:1921tu, Klein:1926tv}, who attempted to combine electromagnetism and gravity in a 5D-gravity framework. In fundamental physics, be it the string theory \cite{Green:1987sp}, which is a work in progress in unifying the quantum theory and gravity or in the context of symmetries of particle physics  \cite{Furey:2015yxg, Baez:2001dm, Baez:2010ye, Furey:2018yyy, Furey:2018drh, Gillard:2019ygk}, the extra  dimensions seems to appear {\em naturally}. Development of string theory also motivated the so-called brane-world scenarios-- where our 4D Universe (3-brane) is embedded in a higher dimensional bulk. The Dvali-Gabadadze-Porrati (DGP) models produce infra-red modification with extra dimensional gravity dominating at low energy scale \cite{Dvali:2000hr}. Further generalization to cosmology of DGP model lead to late-accelerated-expansion \cite{Deffayet:2000uy}. 
Perhaps, the most popular of these models are the `warped braneworld' models \cite{Rubakov:1983bb, Gogberashvili:1998iu, Gogberashvili:1998vx,Randall:1999ee, Randall:1999vf} that generate ultra-violet modification to general relativity with extra dimensional gravity dominating at high energy scale. This model proposes a non-factorizable geometry- a curved five-dimensional spacetime in which the 4D-metric is warped by the extra dimension. Though some recent research on wormholes embedded in higher-dimensional spacetime has been published \citep{Lobo:2007qi, deLeon:2009pu, Wong:2011pt, Kar:2015lma, Banerjee:2019ssy, Wang:2017cnd}, warped braneworld models have not been examined.

We showed in \cite{Sharma:2021kqb} that a GEB (and EB) spacetime embedded in 5D warped bulk (5D-WWEB) model meets the energy conditions. As a follow-up, we analysed the timelike trajectories in these spacetimes in detail in \cite{Sharma:2022tiv}. In this work, we investigate the congruence of geodesics in the original 4D-GEB geometry as well as the 5D-WEB spacetime and compare them to see how the wormhole parameter and warped extra dimension effect the congruence evolution. Note that the 5D line element we utilised is the well-known thick brane model \cite{Dzhunushaliev:2009va, Koley:2004at,Zhang:2007ii,Ghosh:2008vc}, in which the warp factor is a smooth function of the extra dimension (thus there are no derivative jump or delta functions in the curvature and connections). 

The following is a breakdown of the structure of this article. Section (\ref{sec:2}) introduces the wormhole spacetimes that correspond to the novel 5D-WGEB model. It includes a summary of the geometric properties and geodesic equations. In Section (\ref{sec:3}), we used timelike velocity vector field to obtain analytic expressions for the expansion and shear (with zero rotation) variables corresponding to the geodesic congruences and solve the Raychaudhury equations for 4D-GEB geometry.  In section (\ref{sec:4}), we numerically computed the ESR variables for two cases-- without rotation and with rotation for both 4D-GEB and 5D-WGEB geometries. In Section (\ref{sec:5}), to get further insight about the congruences in 4D and 5D, the evolution of cross-sectional area of a geodesic congruence of timelike geodesics is numerically determined and presented graphically. Finally, in Section (\ref{sec:6}) we summarise the key results and discuss future direction of research.


\section{The 5D-WGEB geometry and Geodesics} \label{sec:2}

The specific line element of 5D-WGEB geometry introduced in \cite{Sharma:2021kqb} is as follows,
\begin{equation}
ds^{2} =  e^{2f(y)} \Big[ - dt^{2} +  dl^{2} + r^{2}(l)~\big(  d\theta^{2} + \sin^{2}\theta~d\phi^{2} \big) \Big] + dy^{2}, \label{eq:5d-line-element}
\end{equation}
where, the extra spatial dimension is represented by $y$ ($ - \infty \leq y \leq \infty$). $f(y)$ is the warping factor that we chose as, $f(y) = \pm \log[\cosh(y/y_{0})]$, corresponding to a thick brane scenario \cite{Dzhunushaliev:2009va}. The growing warp factor solution was found using tachyon field in \cite{Koley:2004at} and the decaying warp factor solution was found using phantom scalar field in \cite{Zhang:2007ii}. 5D bulk with a cosmological thick brane in presence of both decaying and growing warp factor is constructed numerically in \cite{Ghosh:2008vc}. The four dimensional part of the above metric (\ref{eq:5d-line-element}) is a spherically symmetric, ultra-static Generalized Ellis-Bronnikov wormhole spacetime, given by,
\begin{equation}
ds^{2}_{4D} = - dt^{2} +  dl^{2} + r^{2}(l)~\big(  d\theta^{2} + \sin^{2}\theta~d\phi^{2} \big) \label{eq:generalised-E&B-l}
\end{equation}
\begin{equation}
\mbox{where}~~~~~ r(l) = (b_{0}^{m} + l^{m})^{1/m}. \label{eq:r(l)}
\end{equation}
Here, $l$ is the `proper radial distance' or the `tortoise coordinate'. The `radius' of the wormhole throat is given by $b_{0}$, and $m$ is the wormhole parameter that essentially generalises the EB geometry ($m=2$) to GEB geometry($m > 2$). For smoothness of $r(l)$, it is necessary to have even valued $m$. In the usual radial coordinate $r$, the above metric (\ref{eq:generalised-E&B-l}) can be written as,
\begin{equation}
ds^{2} = - dt^{2} + \frac{dr^{2}}{\Big( 1 - \frac{b(r)}{r} \Big)} + r^{2} \big( d\theta^{2} + \sin^{2}\theta d\phi^{2} \big), \label{eq:generalised-E&B-r}
\end{equation}
where $r$ and $l$ are related through the {\em shape function} $b(r)$ as,
\begin{equation}
dl^{2} = \frac{dr^{2}}{ \Big( 1 - \frac{b(r)}{r} \Big)}~~~~~\implies ~~~~~\quad b(r) = r - r^{(3-2m)} (r^{m} - b_{0}^{m})^{ \Big( 2 - \frac{2}{m} \Big)}. \label{eq:r-l-relation}
\end{equation}
The geometric and curvature quantities for both the 5D-WGEB and 4D-GEB spacetimes have been discussed in detail in \cite{Sharma:2021kqb}. Interested reader is requested to go through that article. In \cite{Sharma:2022tiv}, we discussed the timelike geodesics in detail. In the following we shall focus on the congruence of geodesics. 
Below we rewrite the geodesic equations corresponding to metrics (\ref{eq:5d-line-element}) and (\ref{eq:generalised-E&B-l}) which are to be solved along with the geodesic deviation equations to determine the properties of the congruences.
The geodesic equations for 4D-GEB are,
\begin{equation}
\frac{d^{2}t}{d\lambda^{2}} = 0 \label{eq:geodesic-1}
\end{equation}
\begin{equation}
\frac{d^{2}l}{d\lambda^{2}} - l^{-1+m}  ~\big( b_{0}^{m} + l^{m} \big)^{-1 + \frac{2}{m}} ~\Big[ \Big( \frac{d\theta}{d\lambda} \Big)^{2} + \sin^{2}\theta  ~\Big( \frac{d\phi}{d\lambda} \Big)^{2} \Big] = 0 \label{eq:geodesic-2}
\end{equation}
\begin{equation}
\frac{d^{2}\theta}{d\lambda^{2}} + \frac{2l^{-1+m}}{\big( b_{0}^{m} + l^{m} \big)} \frac{dl}{d\lambda} \frac{d\theta}{d\lambda} - \sin\theta \cos\theta \Big( \frac{d\phi}{d\lambda} \Big)^{2} = 0 \label{eq:geodesic-3}
\end{equation}
\begin{equation}
\frac{d^{2}\phi}{d\lambda^{2}} + 2 \cot\theta \frac{d\theta}{d\lambda} \frac{d\phi}{d\lambda} + \frac{2l^{-1+m}}{ \big( b_{0}^{m} + l^{m} \big) } \frac{dl}{d\lambda} \frac{d\phi}{d\lambda} = 0 \label{eq:geodesic-4}
\end{equation}
and the geodesic equations for the 5D-WGEB model are as follows. 
\begin{equation}
\frac{d^{2}t}{d\lambda^{2}} + 2 ~f'(y) ~\frac{dt}{d\lambda} ~\frac{dy}{d\lambda} = 0 \label{eq:geodesic-11}
\end{equation}
\begin{equation}
\frac{d^{2}l}{d\lambda^{2}} + 2~f'(y)~\frac{dl}{d\lambda}~\frac{dy}{d\lambda}  - l^{-1+m}  ~\big( b_{0}^{m} + l^{m} \big)^{-1 + \frac{2}{m}} ~\Big[ \Big( \frac{d\theta}{d\lambda} \Big)^{2} + \sin^{2}\theta  ~\Big( \frac{d\phi}{d\lambda} \Big)^{2} \Big] = 0 \label{eq:geodesic-22}
\end{equation}
\begin{equation}
\frac{d^{2}\theta}{d\lambda^{2}} + 2 ~f'(y) ~\frac{d\theta}{d\lambda} ~\frac{dy}{d\lambda} + ~\frac{2l^{-1+m}}{(b_{0}^{m} + l^{m})} ~\frac{d\theta}{d\lambda} ~\frac{dl}{d\lambda} - \sin\theta ~\cos\theta ~\Big( \frac{d\phi}{d\lambda} \Big)^{2} = 0 \label{eq:geodesic-33} 
\end{equation}
\begin{equation}
\frac{d^{2}\phi}{d\lambda^{2}} + 2 ~f'(y) ~\frac{d\phi}{d\lambda} ~\frac{dy}{d\lambda} + 2 ~\cot\theta ~\frac{d\theta}{d\lambda} ~\frac{d\phi}{d\lambda} + \frac{2l^{-1+m}}{ \big( b_{0}^{m} + l^{m} \big) } ~\frac{dl}{d\lambda} ~\frac{d\phi}{d\lambda} = 0 \label{eq:geodesic-44}
\end{equation}
\begin{equation}
\frac{d^{2}y}{d\lambda^{2}} + f'(y)~e^{2f(y)}~\Big[ \Big( \frac{dt}{d\lambda} \Big)^{2} - \Big( \frac{dl}{d\lambda} \Big)^{2} - (b_{0}^{m} + l^{m})^{2/m} ~\Big[ \Big( \frac{d\theta}{d\lambda} \Big)^{2} + \sin^{2}\theta ~\Big( \frac{d\phi}{d\lambda} \Big)^{2} \Big] \Big] = 0 \label{eq:geodesic-55}
\end{equation}
 Here $\l$ is the affine parameter. The differences between 4D and 5D geodesic equations can be seen explicitly from the above expressions, which are $y$ and $\dot{y}$ dependent term in the right hand sides of the Eqs. (\ref{eq:geodesic-11}) to (\ref{eq:geodesic-55}), and an extra equation for motion along the extra dimension. Note that Eq. \ref{eq:geodesic-55}, implies that in presence of growing warp factor we have confined trajectories (whose  $y$-coordinate oscillate about $y=0$) and in presence of a decaying warp factor we have runaway trajectories ($y \ra \pm \infty$ with evolving $\l$) \cite{Sharma:2022tiv,Ghosh:2009ig}.

\section{Derivation of ESR From the Velocity Field} \label{sec:3}

\begin{figure}[H]
\centering
\includegraphics[scale=.5]{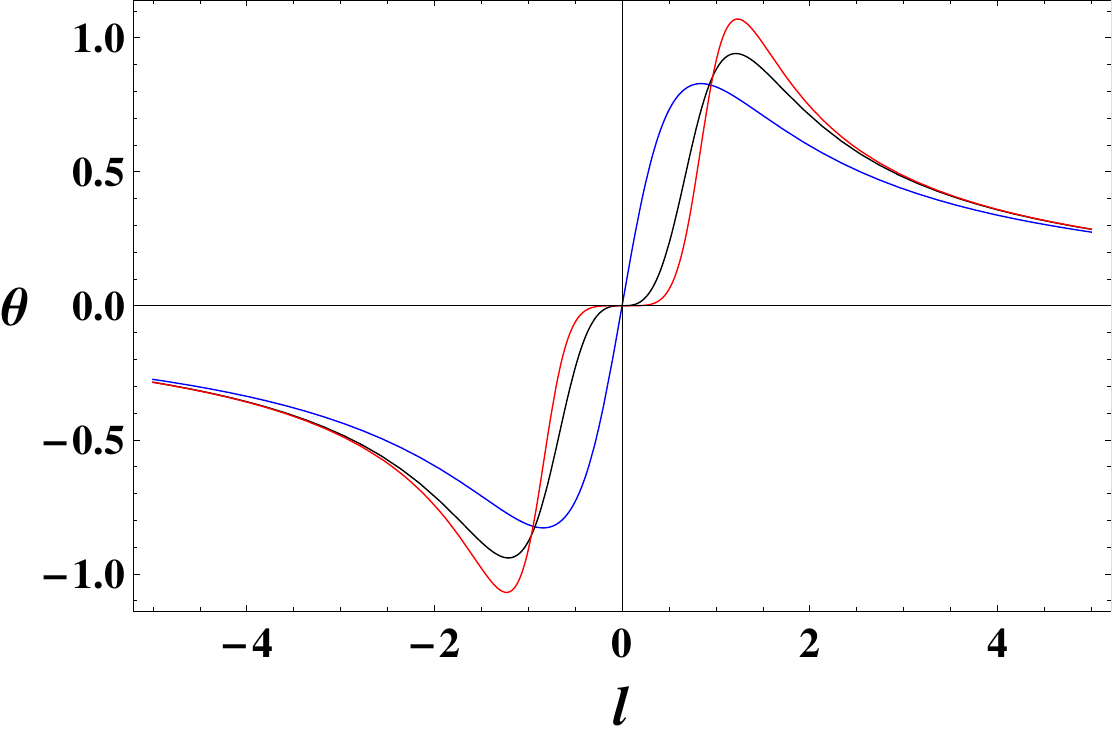}
\hspace{1cm}
\includegraphics[scale=.5]{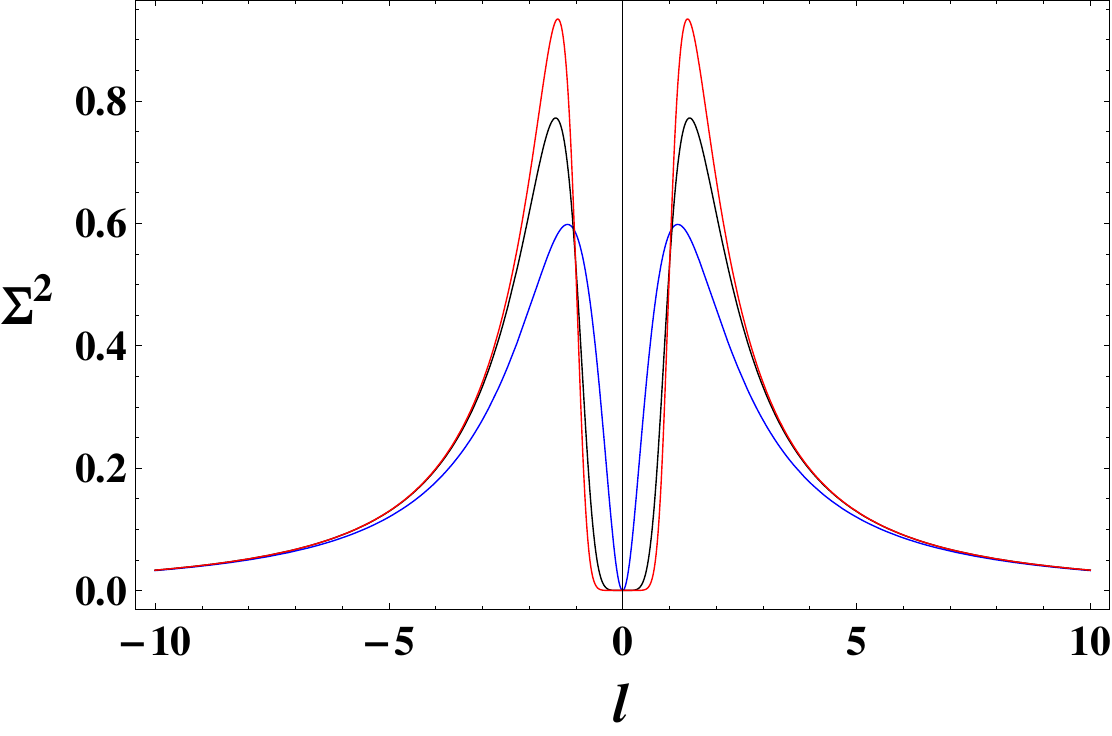}
\caption{Evolution of $\Theta$ and $\Sigma^{2}$ in case of crossing geodesic congruences with different choice of wormhole parameter $m = 2, 4, 6$ (blue , black and red curves). Where, $b_{0} = 1$, $k = \sqrt{3}$ and $h = 1$. }
\label{fig:GEB-ESR-analytic}
\end{figure}

In the following we solve the the Raychaudhuri equations to analyse the flow of geodesic congruences through the kinematic variables expansion $\Theta$, shear $\Sigma_{AB}$ and rotation $\Omega_{AB}$ (ESR). Note that we shall consider those trajectories that cross the throat (for details of {\em crossing trajectories} see \cite{Sharma:2022tiv}). 
The Raychaudhuri equation for expansion \cite{Kar:2006ms,Poisson:2009pwt}, given by  
\begin{equation}
\frac{d\Theta}{d\lambda} + \frac{1}{3}\Theta^{2} + R_{AB}u^{A}u^{B} + \Sigma^{2} - \Omega^{2} = 0, \label{eq:raychaudhuri-eq}
\end{equation}
is dependent on other kinematic variables such as the shear and rotation, as well as the curvature term, $R_{AB}u^{A}u^{B}$. The above equation is a non-linear first order differential equation that can be transformed into a second order {\em linear} form via redefining the $\Theta$ as $\Theta = 3 \frac{\dot{F}}{F} $. Thus, Eq. (\ref{eq:raychaudhuri-eq}) turn out to be,
\begin{equation}
\frac{d^{2}F}{d\lambda^{2}} + \frac{1}{3} \left( R_{AB}u^{A}u^{B} + \Sigma^{2} - \Omega^{2} \right)F = 0. \label{eq:raychaudhuri-eq-2}
\end{equation}
The congruence converges at a finite $\lambda$ where $F = 0$, $\dot{F} < 0$. Using the Sturm comparison theorems in differential equations, one can show that convergence happens when 
\begin{equation}
R_{AB}u^{A}u^{B} + \Sigma^{2} - \Omega^{2} \geq 0 .\label{eq:convergence-condition}
\end{equation}
The role of the terms that appear in the evolution equation for expansion, is clearly shown in the above convergence condition. While rotation works against convergence, shear works in its favour. When $\Omega_{AB} = 0$ then $R_{AB}u^{A}u^{B} \geq 0$ leads to focusing of the congruences. 
In the case of vanishing rotation, we derived analytic expressions for ESR variables using the first integral of the metric and plotted them below.
ESR variables can be derived directly from their definitions where the velocity vector field is $u^{A}$. The following are the formal definitions for the expansion $\Theta$, shear $\Sigma_{AB}$, and rotation $\Omega_{AB}$ \citep{Poisson:2009pwt}:
\begin{equation}
\Theta = \nabla_{A}u^{A} , \label{eq:theta-definition}
\end{equation}
\begin{equation}
\Sigma_{AB} = \frac{1}{2}~\left( \nabla_{B}u_{A} + \nabla_{A}u_{B}  \right) - \frac{1}{n - 1}~h_{AB}~\Theta , \label{eq:sigma-definition}
\end{equation}
\begin{equation}
\Omega_{AB} = \frac{1}{2}~\left( \nabla_{B}u_{A} - \nabla_{A}u_{B}  \right). \label{eq:omega-definition}
\end{equation}
Here, $n$ is the dimension of spacetime, $h_{AB} = g_{AB} \pm u_{A}u_{B} $ is the projection tensor and $u_{A}u^{A} = \mp 1$ (The plus and minus sign stands for the timelike and spacelike geodesics, respectively). In case of the 4D-GEB geometry the calculation of the first integral $u^{A}$ (whose expressions are independent of the affine parameter $\lambda$) corresponding to each coordinate could be found. These are listed bellow in the case of timelike trajectories (for sake of simplicity, we choose $\theta = \pi/2 $ without loosing any generality).
\begin{equation}
\dot{t} = k ,\label{eq:t-dot}
\end{equation}
\begin{equation}
\dot{\phi} = \frac{h}{\left( b_{0}^{m} + l^{m} \right)^{\frac{2}{m}}} ,\label{eq:phi-dot}
\end{equation}
\begin{equation}
\dot{l} = \sqrt{k^{2} - \frac{h^{2}}{\left( b_{0}^{m} + l^{m} \right)^{\frac{2}{m}}} - 1} .\label{eq:l-dot}
\end{equation}
Eq. (\ref{eq:l-dot}) obtained by using Eqs, (\ref{eq:t-dot}), (\ref{eq:phi-dot}) and timelike constraint on geodesics, $ g_{AB}u^{A}u^{B} =  -1  $. Here, $k$ and $h$ are the conserved energy and angular momentum per unit mass of the timelike particle. 
For given velocity vector field (Eqs. \ref{eq:t-dot} to \ref{eq:l-dot}) of 4D-GEB wormhole geometry, the expansion scalar and amplitude squared of shear is,
\begin{equation}
\theta = l^{-1+m} \left[ \frac{h^{2} + \left(b_{0}^{m} + l^{m} \right)^{\frac{2}{m}} \left(k^{2} - \frac{h^{2}}{\left(b_{0}^{m} + l^{m} \right)^{\frac{2}{m}}} - 1 \right)}{\left(b_{0}^{m} + l^{m} \right)^{1 + \frac{2}{m}} \sqrt{k^{2} - \frac{h^{2}}{\left(b_{0}^{m} + l^{m} \right)^{\frac{2}{m}}} - 1}} \right] , \label{eq:GEB-theta}
\end{equation}
\begin{equation}
\resizebox{\textwidth}{!}
{%
$ \Sigma^{2} = \frac{l^{ -2 + 2m } \left(  b_{0}^{m} + l^{m} \right)^{\frac{-2 \left( 1 + m \right)}{m}} \left[ h^{4} \left( 13 + 4k^{2} + k^{4} \right) - h^{2} \left( - 16 + 12k^{2} + 3k^{4} + k^{6} \right) \left( b_{0}^{m} + l^{m} \right)^{2/m} + \left( - 1 + k^{2} \right)^{2} \left( 9 - k^{2} + k^{4} \right) \left( b_{0}^{m} + l^{m} \right)^{\frac{4}{m}} \right]}{9 \left[ - h^{2} + \left( - 1 + k^{2} \right) \left(  b_{0}^{m} + l^{m} \right)^{\frac{2}{m}} \right]} $%
} \label{eq:GEB-sigsq}
\end{equation}
%
%
One can easily check that if $l \rightarrow 0$ or $l \rightarrow \pm \infty$ the expansion scalar $\Theta \rightarrow 0$ and the shear scalar $\Sigma^{2}\rightarrow$ $0$. This is true irrespective of the value of $h$. Thus there will be no expansion/contraction and distortion of timelike geodesic congruences at the wormhole throat and at asymptotic flat regions which is expected. The profile of $\S^2$ is symmetric about $l=0$, on the other hand, $\th$ profile is exactly asymmetric about the throat. Fig.(\ref{fig:GEB-ESR-analytic}) clarifies the effect of the wormhole parameter $m$ on expansion and shear. The same $\Theta$-variation has also been reported in \cite{DuttaRoy:2019hij}. While plotting the figures the numerical values are chosen to be consistent with the corresponding `crossing trajectory conditions' \cite{Sharma:2022tiv}. 
Variation in  $\Theta$ (or $\Sigma^{2}$) is qualitatively similar in the asymptotic regions for $m = 2$ and $m > 2$ geometries. However, they are very different at about the throat which is a reminiscent of the fact that upto $m^{th}$ order derivative of the geodetic potential vanishes at the throat \cite{Sharma:2022tiv}. Thus a congruence of geodesics coming from $l=\infty$, with zero initial ESR, expands and then contract before crossing the throat with zero ESR. However, the location of the extrema moves farther away from the throat with increasing $m$. 
In case of 5D-WGEB model, it is difficult build any intuition from the equations because of their complexity. Therefore, in the following we numerically analyse the 4D and 5D scenario that is difficult to solve analytically.



\section{Numerical Analysis of ESR variables} \label{sec:5}

\begin{figure}[H]
\centering
\includegraphics[scale=.4]{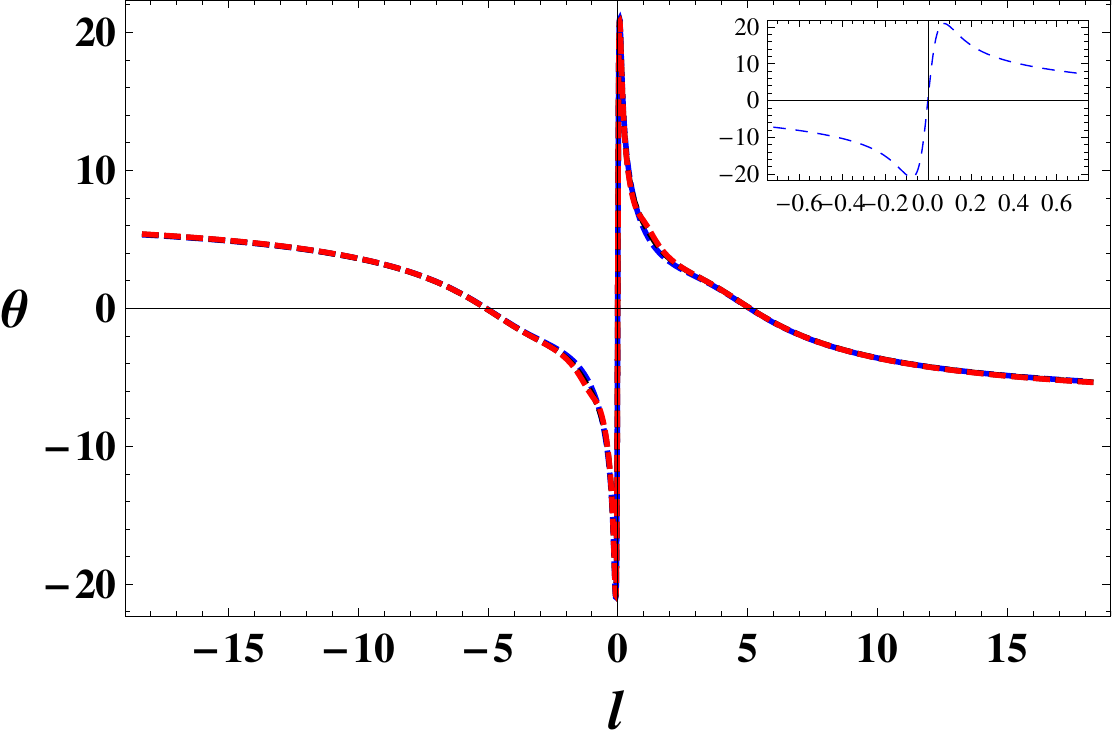}
\hspace{0.2cm}
\includegraphics[scale=.4]{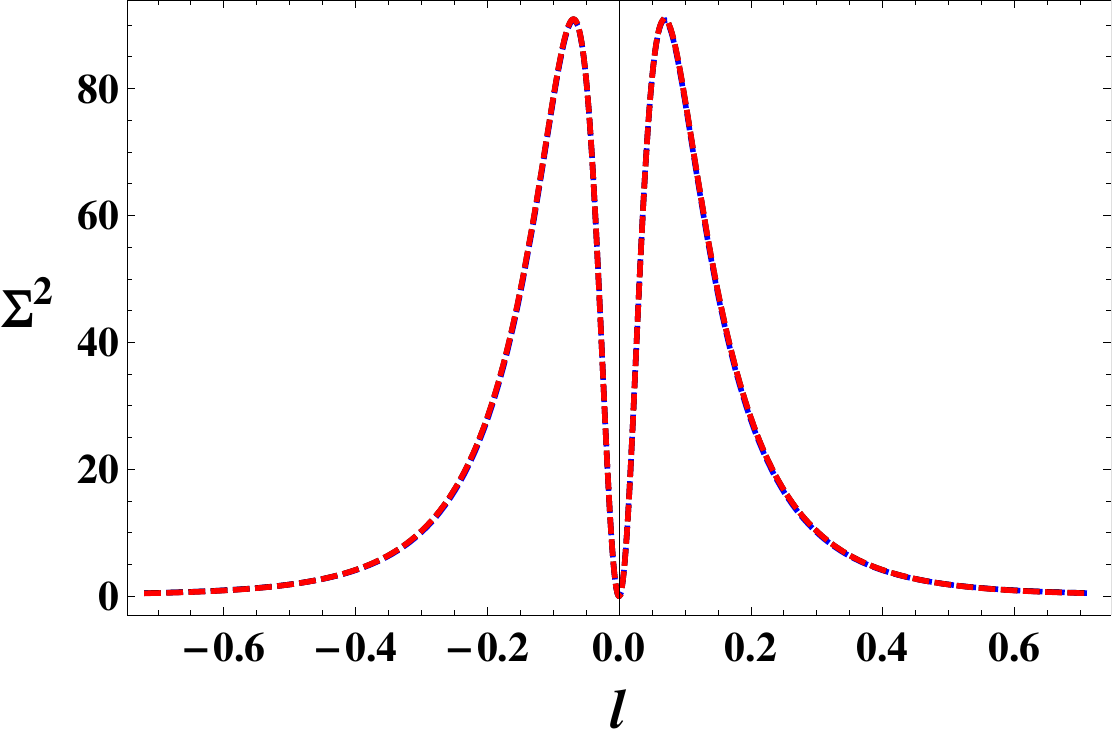}
\hspace{0.2cm}
\includegraphics[scale=.4]{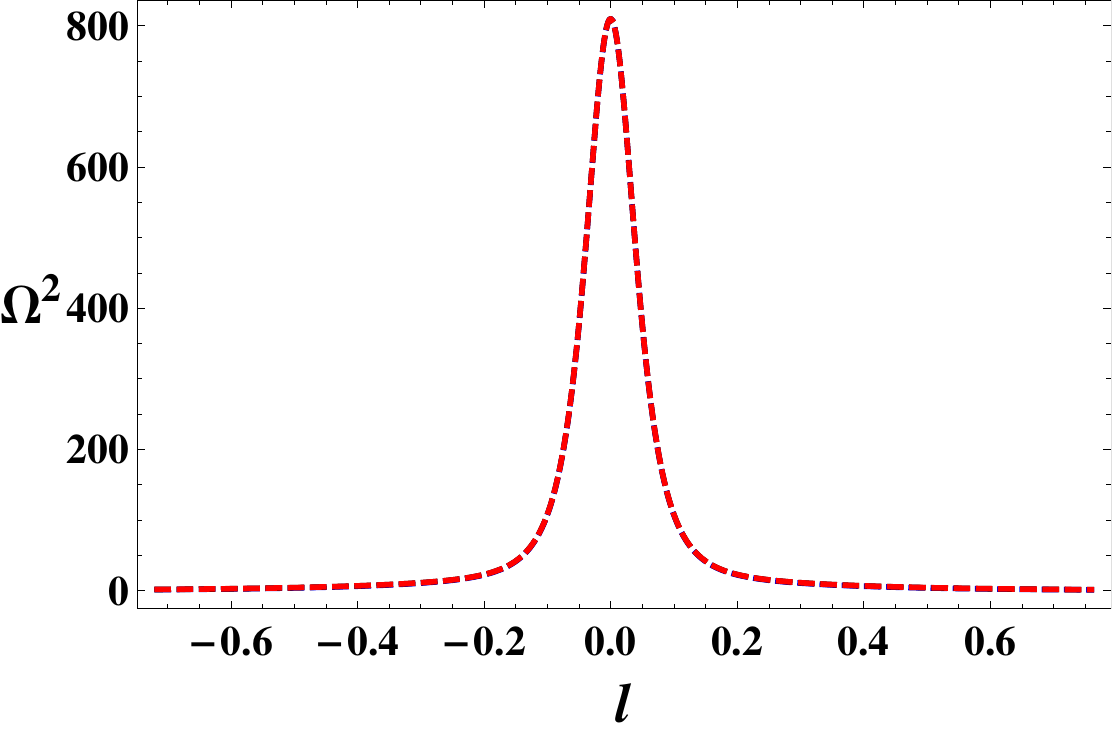}
\hspace{0.2cm}
\caption{Showing evolution of ESR  in case of crossing geodesic congruences  for decaying warp factor wormhole parameter, $m = 2, 4$, and $6$ (blue, black and red curves) with as $b_{0} = 1$. }
\label{fig:5d-decay-ESR-wr2}
\end{figure}

To describe the behaviour of evolution of expansion $\Theta$ in both the 4D-GEB and 5D-WGEB models, we numerically solved Eq. (\ref{eq:evolution-of-B_AB}) along with the geodesic equations. For this, we have chosen two types of boundary conditions along with $\Theta$ and $\Sigma_{AB}$ being zero at the throat: (a) without rotation ($\Omega_{AB}=0$) and (b) with rotation ($\Omega_{AB}\neq 0$). The velocity components for timelike geodesics in the 4D-GEB background satisfying the timelike constraint are chosen as $\{ \dot{t}(0) = \sqrt{3} $, $\dot{l}(0) = 1.41421$, $\dot{\theta}(0) = 0$, $\dot{\phi}(0) = 0 \} $  and the same for the 5D-WGEB models with growing and decaying warp factor are  $\{ \dot{t}(0) = 1.71485 $, $\dot{l}(0) = 1.39665 $, $\dot{\theta}(0) = 0$, $\dot{\phi}(0) = 0 $, $ \dot{y(0)} = 0 \}$ and $ \{ \dot{t}(0) = 1.74943 $, $\dot{l}(0) = 1.43195 $, $\dot{\theta}(0) = 0$, $\dot{\phi}(0) = 0 $, $ \dot{y(0) = 0} \}$  respectively (see Appendix and \cite{Sharma:2022tiv}). Fig. (\ref{fig:GEB-ESR-wor2}) to (\ref{fig:5d-decay-ESR-wr2}) show the evolution of ESR variables with (continuous curves) and without (dashed curves) rotation for three different values of $m$ for 4D-GEB and 5D-WGEB (with growing and decaying warp factor) models, respectively. 

\subsection{case-1: Congruences in 4D-GEB spacetime with and without rotation}

The evolution of numerically solved expansion $\Theta$ (without rotation) in the 4D-GEB  scenario  shows exact similarity with the analytic behaviour discussed earlier thus proving the accuracy of the numerical computation. The quantitative difference wrt Fig. \ref{fig:GEB-ESR-analytic} is because we have chosen angular momenta $h$ of the geodesics of the congruence to be zero for computational simplicity. For readers sake, we also showed the evolution of the curvature term $R_{AB}u^{A}u^{B}$ along with the expansion and shear.
The boundary values used in this subsection is shown in Appendix \ref{app-1}. 

%


\begin{figure}[H]
\centering
\includegraphics[scale=.4]{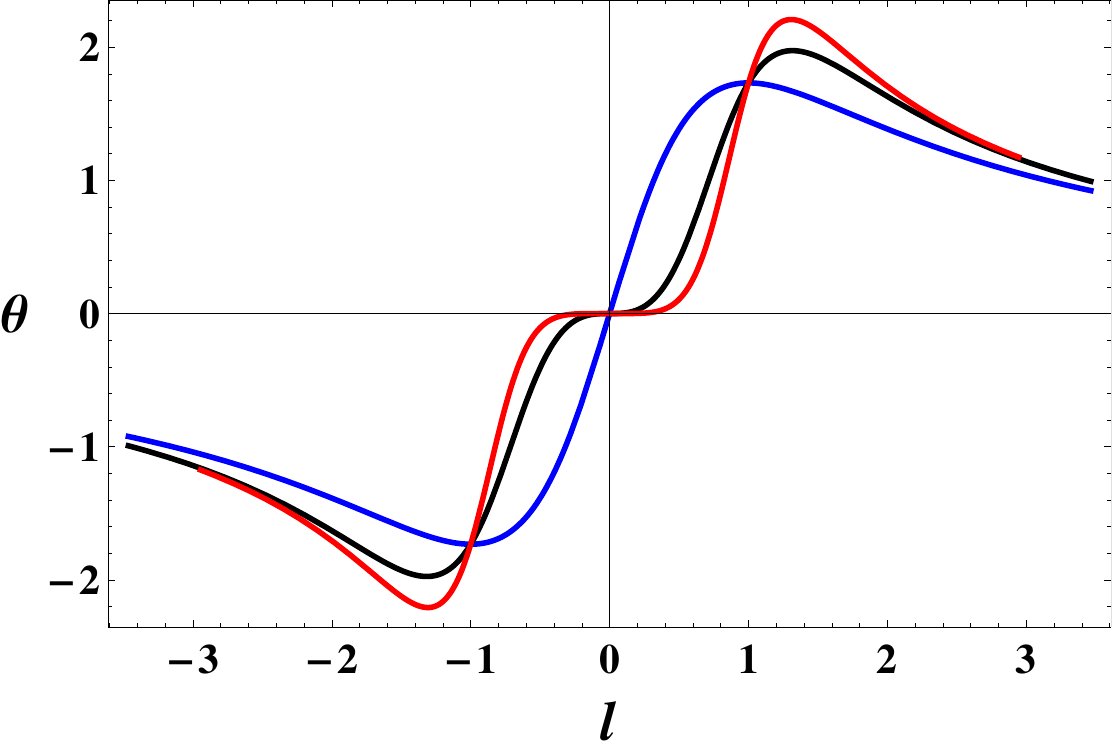}
\hspace{.3cm}
\includegraphics[scale=.4]{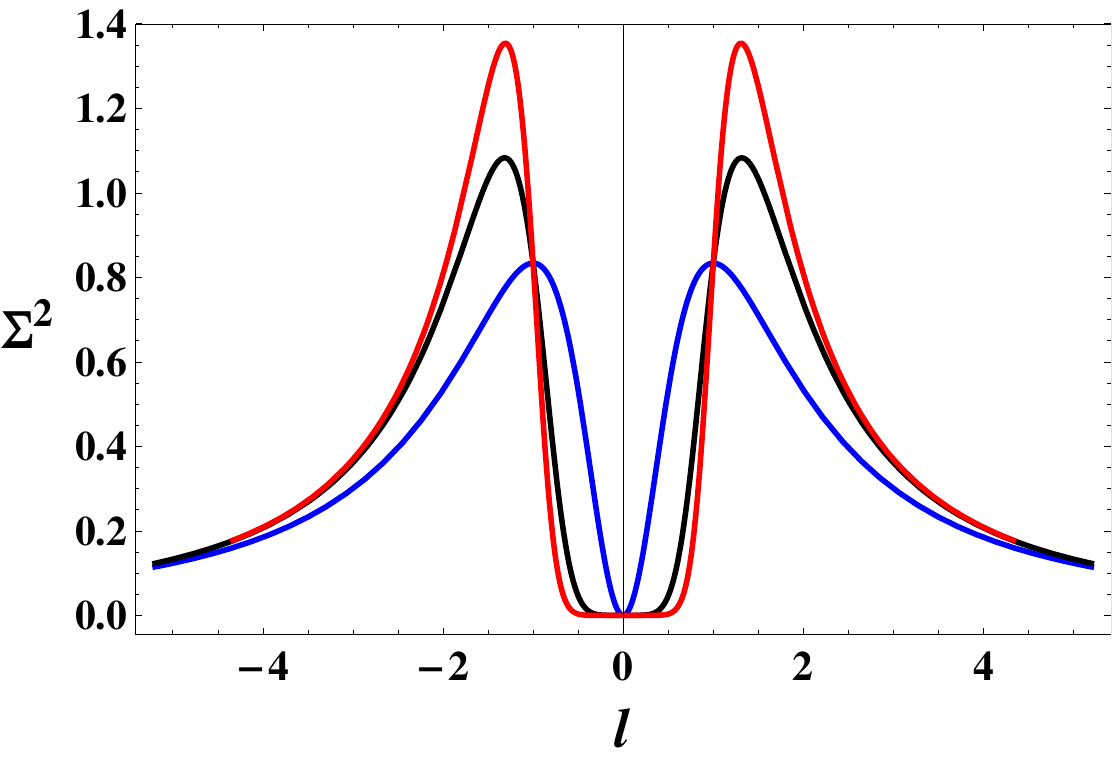}
\hspace{.2cm}
\includegraphics[scale=.45]{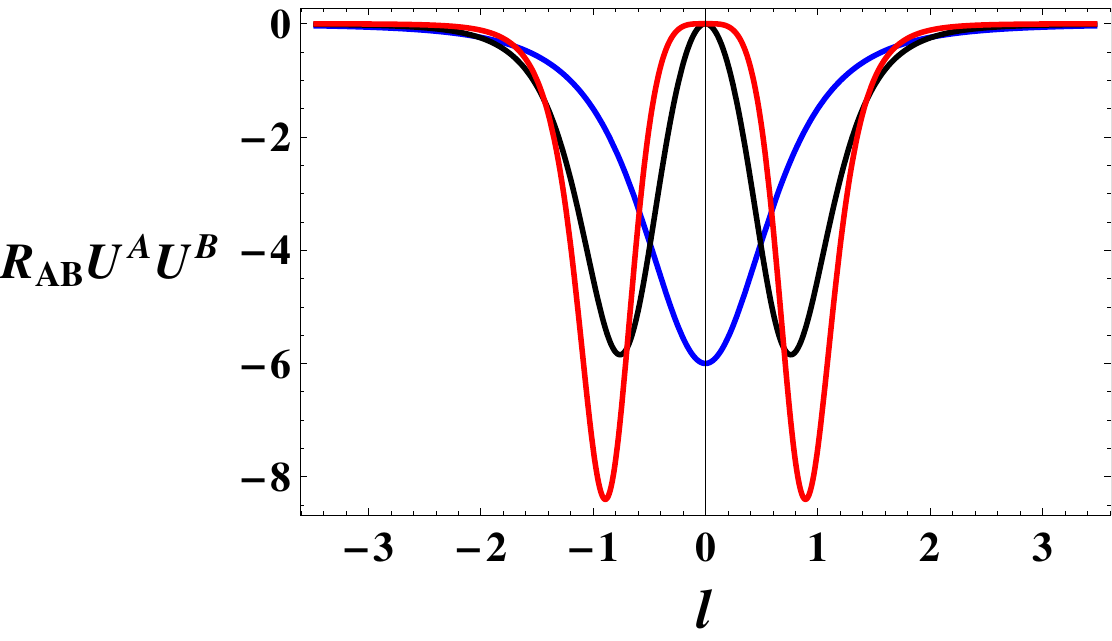}
\caption{Evolution of $\theta$, $\Sigma^{2}$ and the curvature term without rotation in case of crossing geodesic congruences for wormhole parameter, $m = 2, 4$, and $6$ (blue, black and red curves) with $b_{0} = 1$. }
\label{fig:GEB-ESR-wor2}
\end{figure}

\begin{figure}[H]
\centering
\includegraphics[scale=.4]{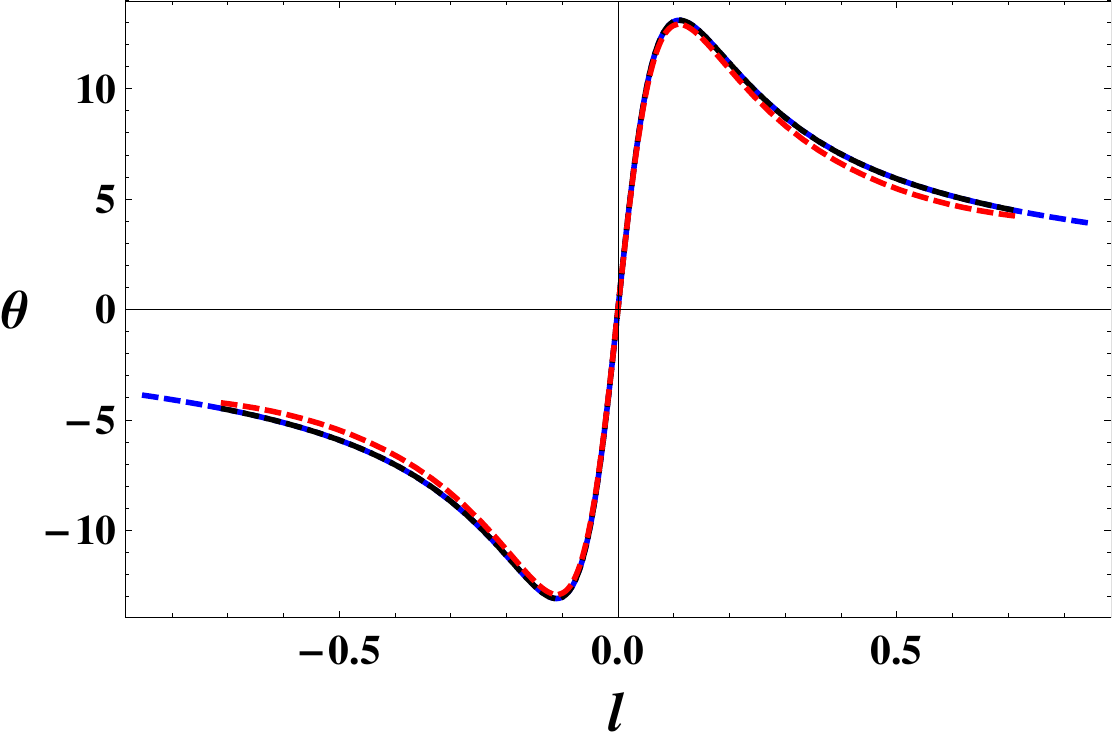}
\hspace{0.2cm}
\includegraphics[scale=.4]{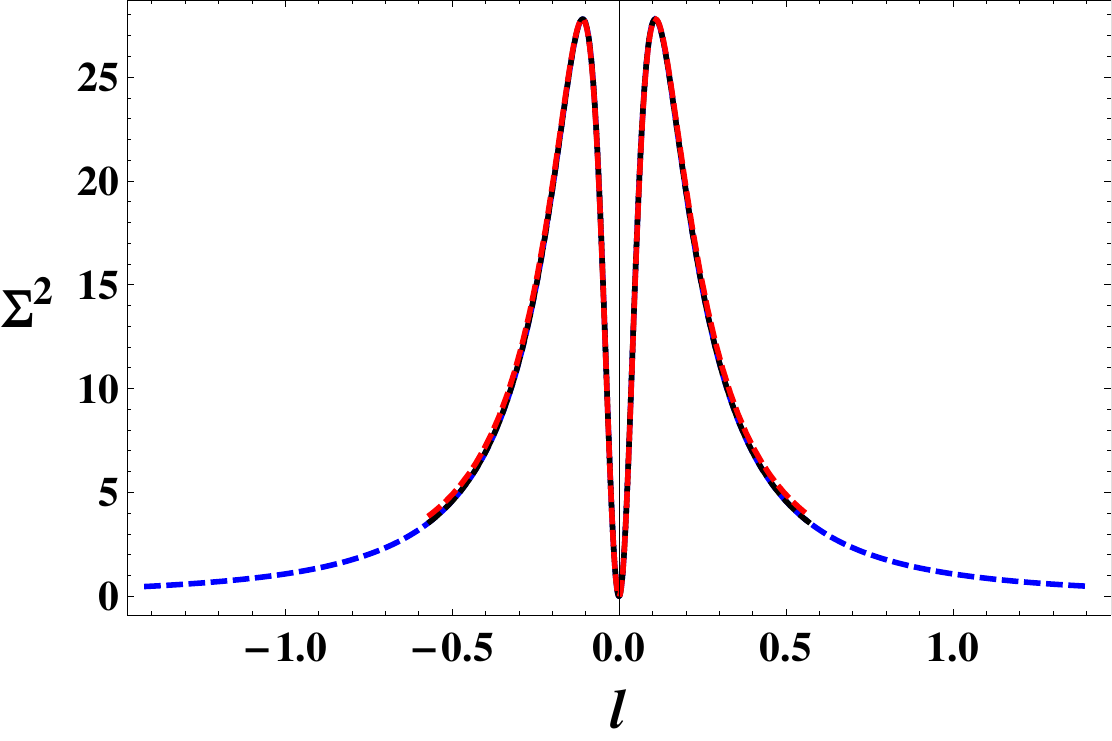}
\hspace{0.2cm}
\includegraphics[scale=.4]{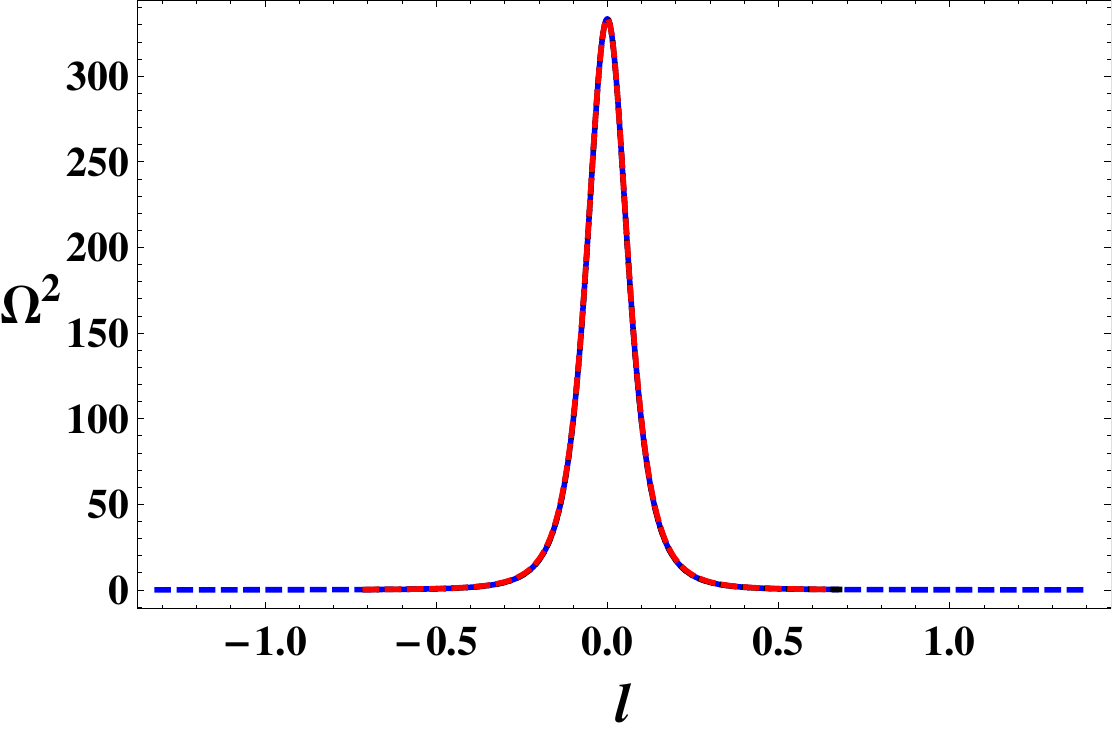}
\hspace{0.2cm}
\caption{Evolution of ESR in case of crossing geodesic congruences with rotation for wormhole parameter, $m = 2, 4$, and $6$ (blue, black and red curves) with $b_{0} = 1$. }
\label{fig:GEB-ESR-wr2}
\end{figure}

Something interesting happens in presence of rotation in the congruence (see Fig. \ref{fig:GEB-ESR-wr2}), even for 4D-GEB that could not be derived through analytic approach. 
Posing a boundary condition at the throat with non-zero rotation $\Omega_{AB}(\l=0) \neq 0$ seems to have an effect of nullifying the effect of $m$ the wormhole parameter. Thus the $\Theta$ profile (vs $l$) for different values of $m$ overlaps. This raises a possibility that a {\em rotating} Ellis-Bronnikov wormhole may not require exotic matter to be stable. However a detailed study of this interesting case is not possible within the scope of this article.  
Fig. \ref{fig:GEB-ESR-wr2} further shows that the initial rotation also dies away in the asymptotic regions.


\subsection{Case-2: Congruences in 5D-WGEB spacetime with growing warp factor}

\begin{figure}[H]
\centering
\includegraphics[scale=.4]{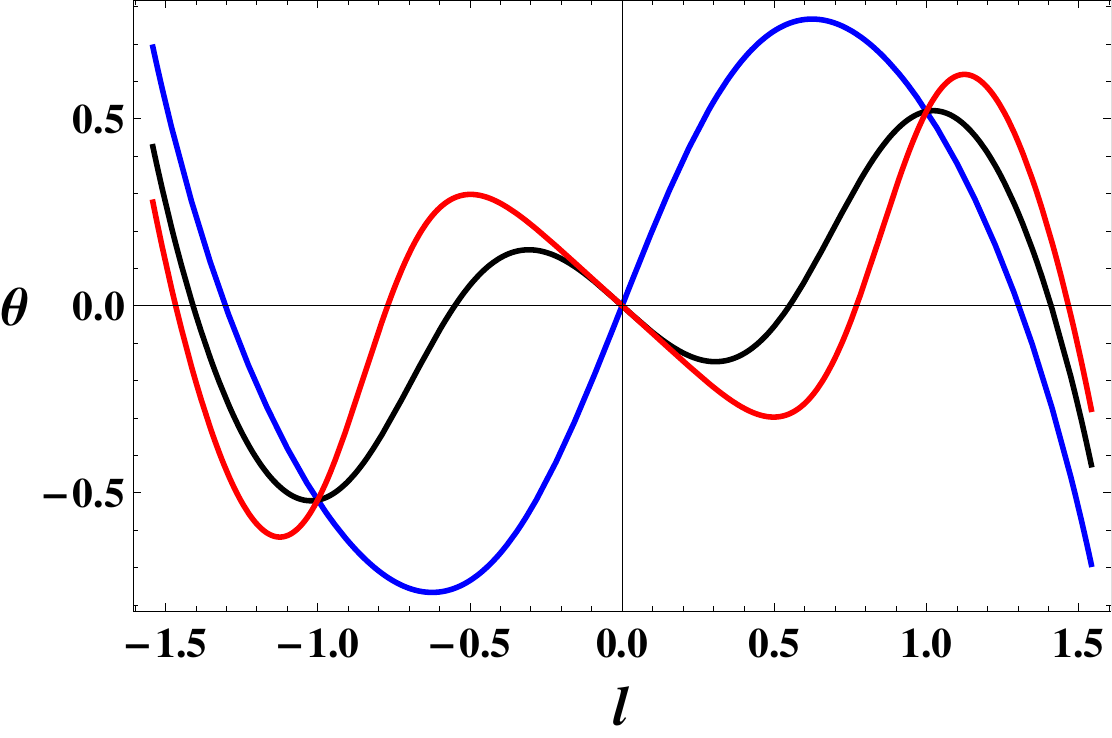}
\hspace{0.2cm}
\includegraphics[scale=.4]{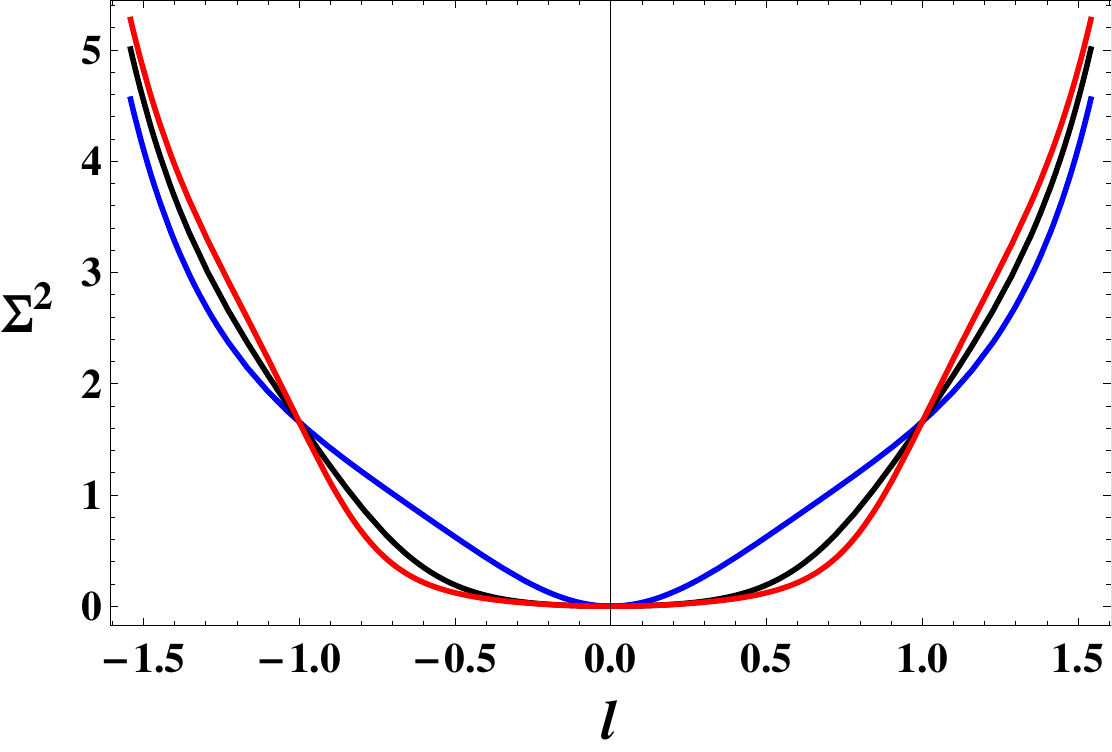}
\hspace{0.2cm}
\includegraphics[scale=.45]{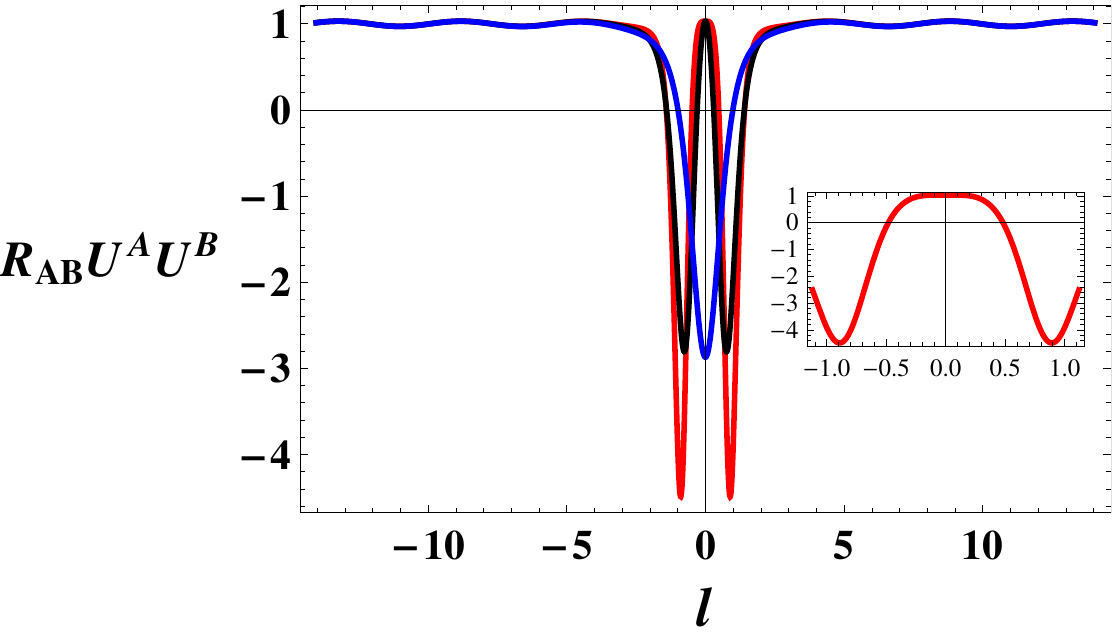}
\caption{Evolution of $\theta$, $\Sigma^{2}$ and the curvature term without rotation in case of crossing geodesic congruences with growing warp factor for $m = 2, 4$, and $6$ (blue, black and red curves) with $b_{0} = 1$. }
\label{fig:5d-grow-ESR-wor2}
\end{figure}

In the presence of growing warp factor, the evolution of ESR variables with rotation and without rotation are shown in figures \ref{fig:5d-grow-ESR-wor2} and \ref{fig:5d-grow-ESR-wr2} respectively.
The boundary values used in this subsection is shown in Appendix \ref{app-2}. 
The general observation on evolution of $\Theta$ implies that a congruence of dense geodesics (focussed) coming from positive $l$ will defocus on the other side and vice versa. The presence of geodesic singularity at finite $l$ is essentially because of bounded trajectories along $y$ \cite{Ghosh:2010gq}. For $m=2$ geometry, there is only one extrema on either side of $l=0$, but for $m > 2$ we have two extrema on either side. 
Shear on the other hand increases monotonically at $l \ra \pm \infty$. The curvature term particularly become positive in the asymptotic region and around the throat which is different from the 4D case where this term was negative everywhere.





Introduction of non-zero rotation, $\Omega_{AB}$, at the throat again compensate for the differences in different $m$ geometries. 
The evolution of ESR variables are similar as the 4D case with rotation to some extent about. Congruence singularity sustains the presence of rotation. Though expansion diverges, shear on the other hand vanishes (unlike the 4D case and without rotation case) in the asymptotic region like rotation. 
\begin{figure}[H]
\centering
\includegraphics[scale=.4]{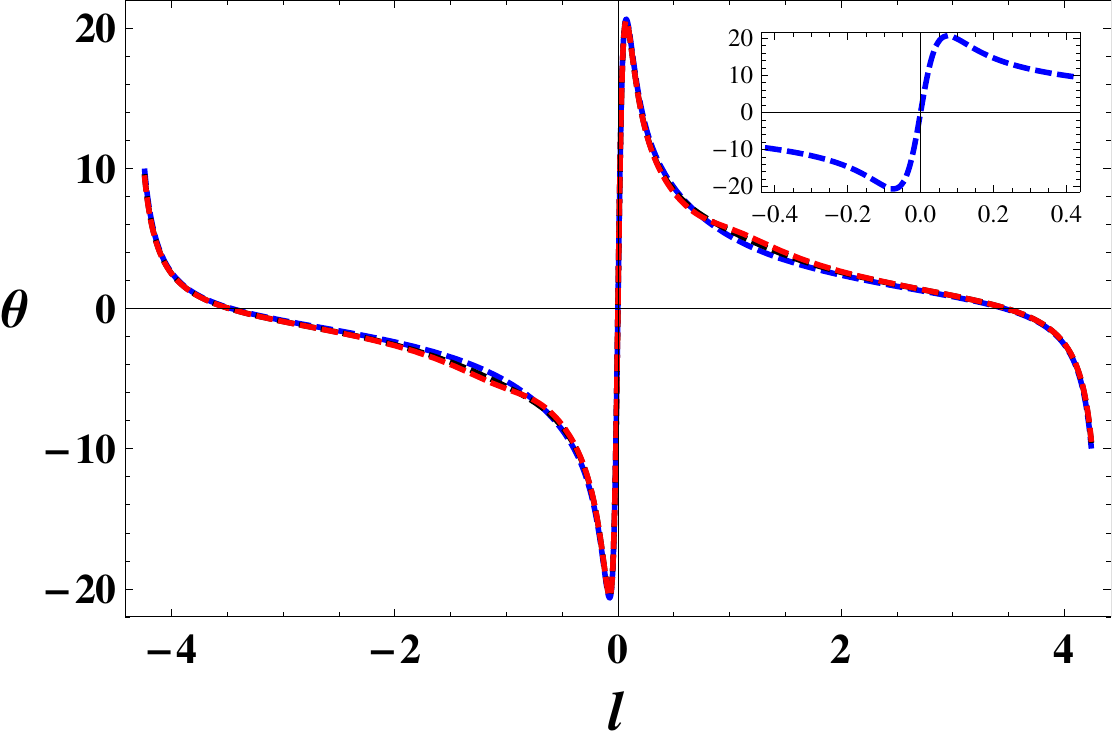}
\hspace{0.2cm}
\includegraphics[scale=.4]{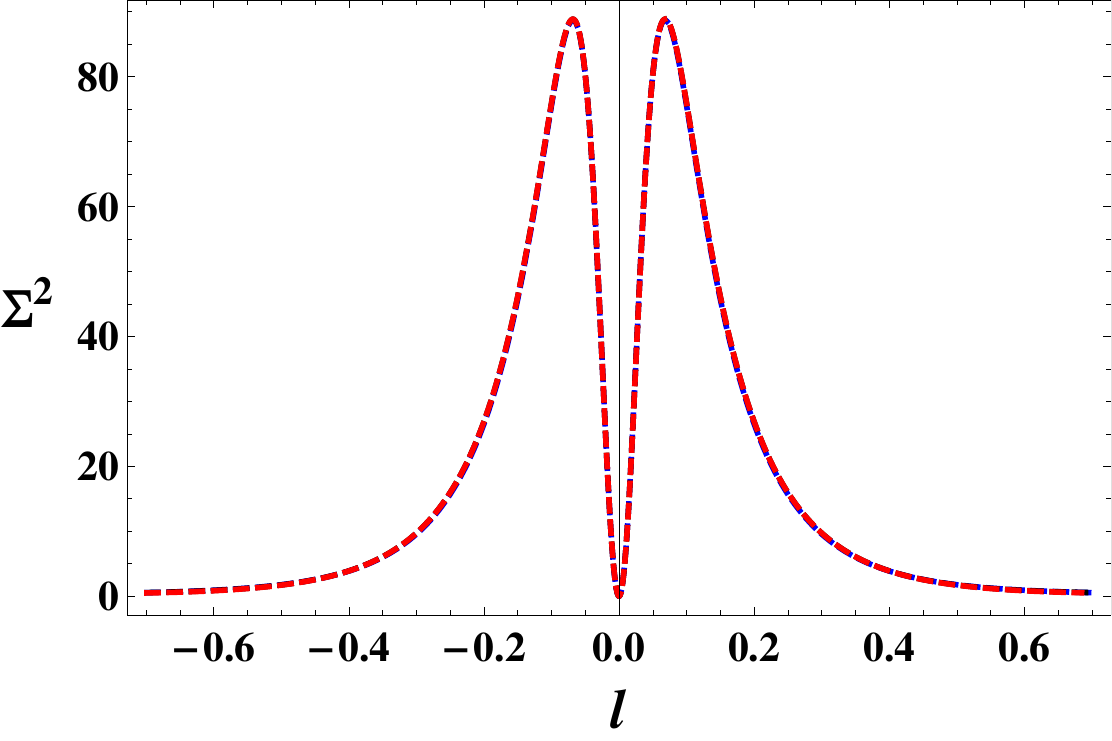}
\hspace{0.2cm}
\includegraphics[scale=.4]{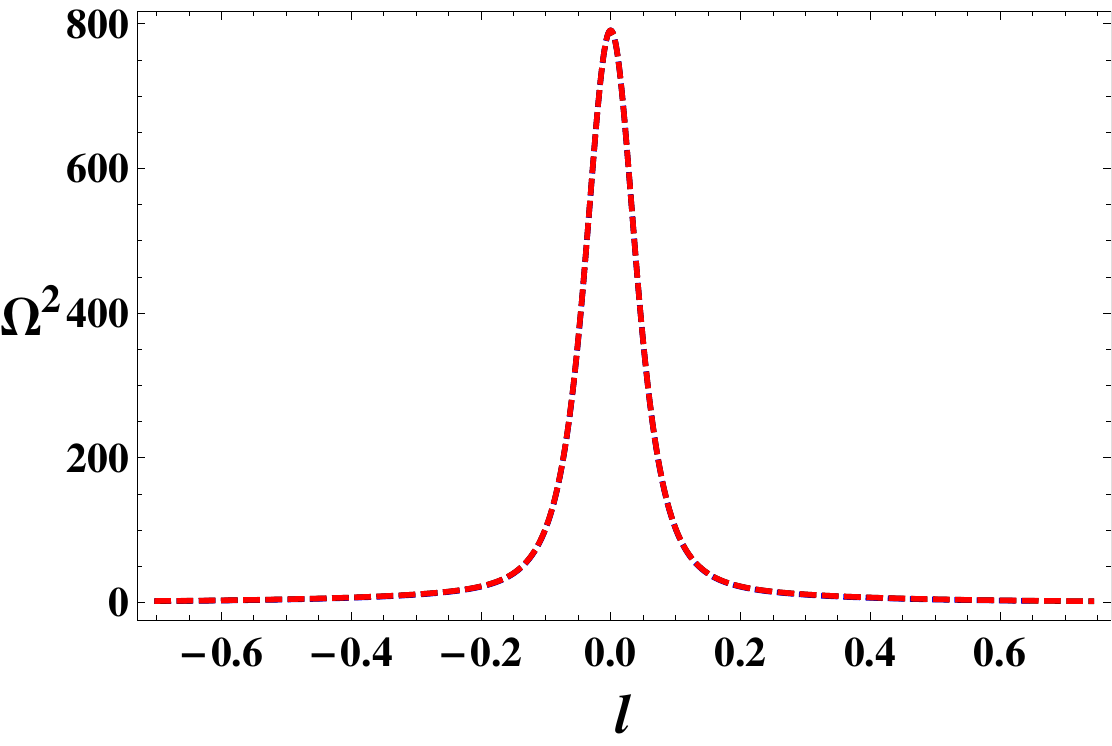}
\hspace{0.2cm}
\caption{Showing evolution of ESR  in case of crossing geodesic congruences  for growing warp factor wormhole parameter, $m = 2, 4$, and $6$ (blue, black and red curves) with as $b_{0} = 1$. }
\label{fig:5d-grow-ESR-wr2}
\end{figure}


\subsection{case-3: Congruences in 5D-WGEB spacetime with decaying warp factor}

\begin{figure}[H]
\centering
\includegraphics[scale=.4]{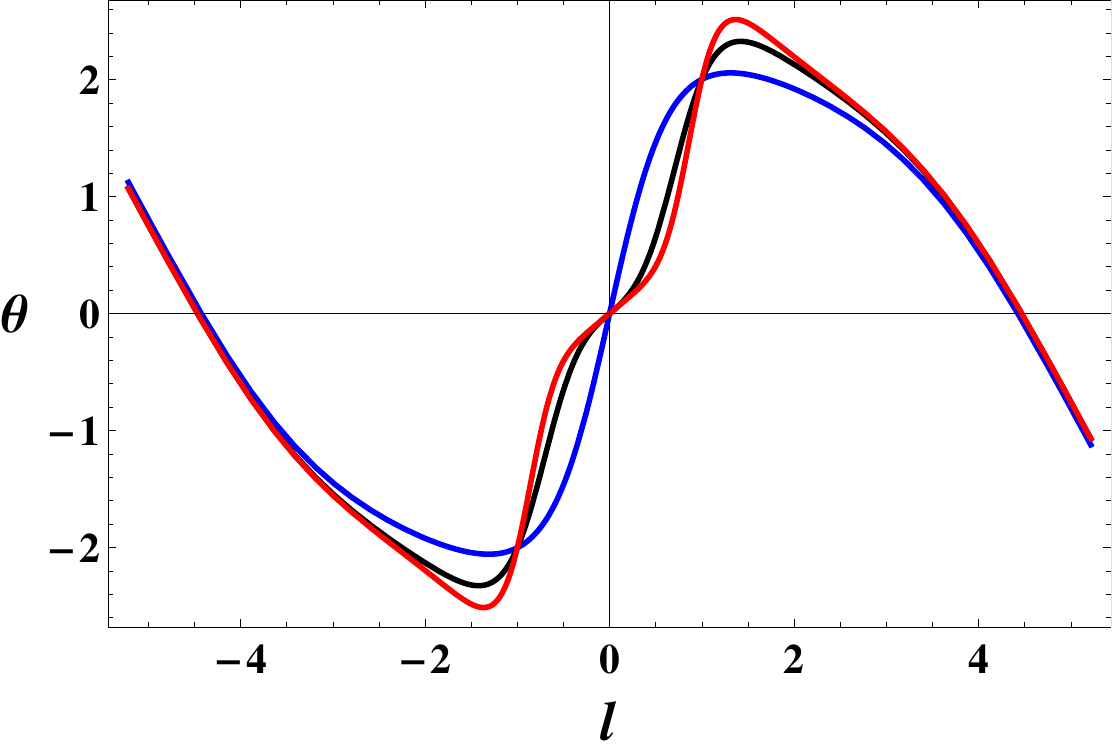}
\hspace{0.2cm}
\includegraphics[scale=.4]{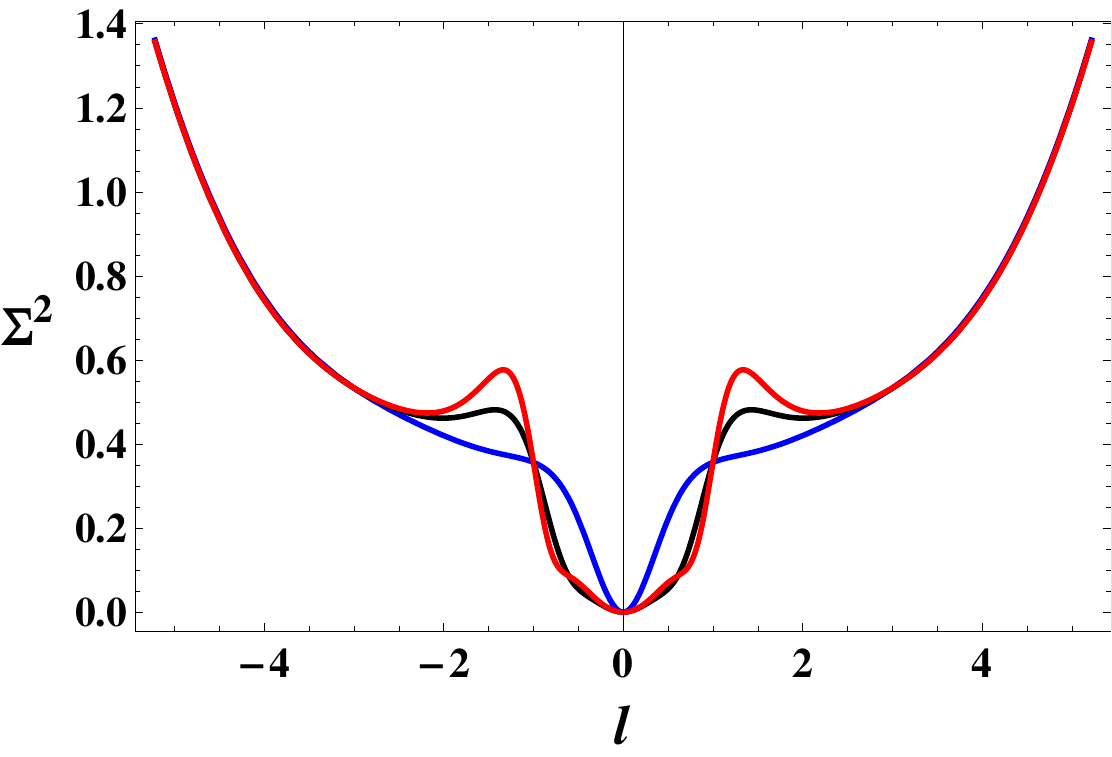}
\hspace{0.2cm}
\includegraphics[scale=.45]{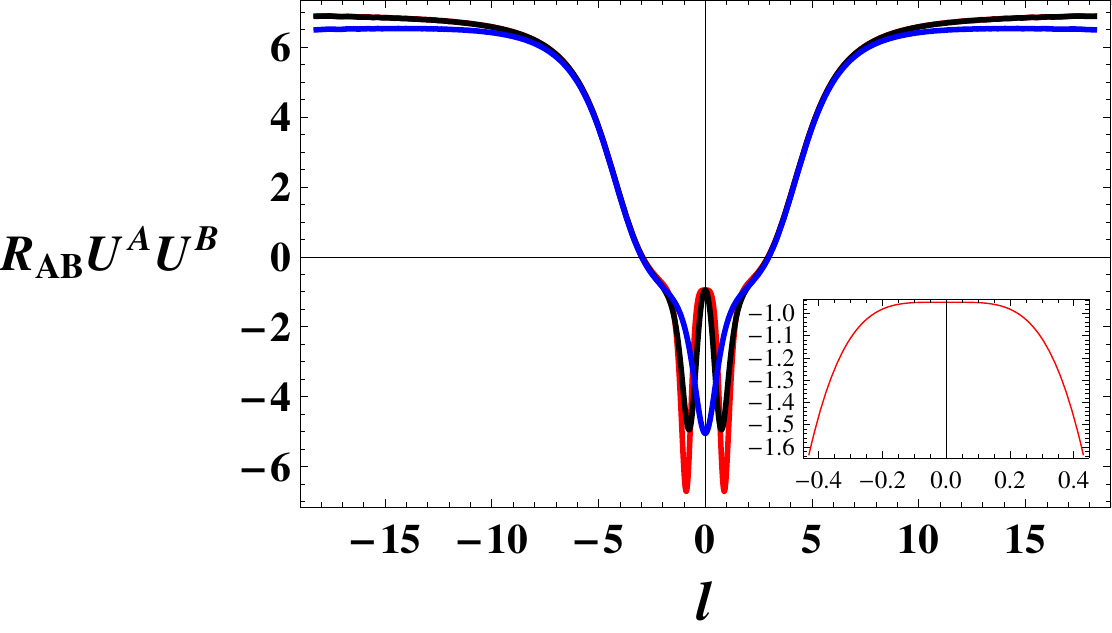}
\caption{Evolution of $\theta$, $\Sigma^{2}$ and the curvature term without rotation in case of crossing geodesic congruences with decaying warp factor for $m = 2, 4$, and $6$ (blue, black and red curves) with $b_{0} = 1$.}
\label{fig:5d-decay-ESR-wor2}
\end{figure}

Fig. \ref{fig:5d-decay-ESR-wor2}, in the presence of a decaying warp factor, without rotation, the differences in the $\Theta$-profiles between $m=2$ and $m > 2$ geometries is decreased-- both have one of extrema on either side of the throat.
Further, the congruence singularities is moved towards $l \ra \pm \infty$.
Shear has local minima at finite $\pm l$ (except the global minima at $l=0$) and increases monotonically beyond them. 
The curvature term, though positive in the asymptotic region, around the throat it contributes negatively (even for $m > 2$) which is similar to the 4D case and opposite to the case with a growing warp factor.


%
%


Introduction of non-zero rotation, $\Omega_{AB}$, at the throat, again compensate for the differences in different $m$ geometries. Remarkably, it slows the evolution down and avoid the divergences at large $l$ as it happened in case with growing warp factor. 
The evolution of shear and rotation are similar to the case with a growing warp factor. 
The boundary values used in this subsection is shown in Appendix \ref{app-3}. 

We can make some general conclusion for the evolution of ESR variables from the figures given above. In case with zero rotation, at $l = 0$ (this is by choice) and at $l = \pm b_{0} = \pm 1$, expansion and shear have same value for all $m$ for both 4D and 5D models. 
The effect of wormhole parameter $m$ neutralises when we introduce rotation ($\Omega_{AB}\neq 0$) in the congruence. Occurrence of divergences in the $\Th$ and $\S^2$ profiles of the geodesic congruences or congruence singularity is due to the extra dimension or the warping factor to be specific. However, the presence of rotation can avoid these divergences in the case of decaying warp factor. 


\section{Evolution of cross-sectional area of Congruences} \label{sec:4}

In this section we present a different perspective on the evolution of ESR variables.
Let us consider a congruence of four timelike geodesics coming from one side ($l = - \infty$) of the throat, crosses the throat, and reaches the other side to infinity ($l = \infty$). One may ask how the cross sectional area of these geodesic congruences evolve as the affine parameter or time evolves? To answer this question, we numerically solve the following set of equations (which are essentially the Raychaudhury and the geodesic deviation equations),
\begin{equation}
u^{C}\nabla_{C}B_{AB} = - B_{AC}B^{C}_{B} - R_{ACBD}u^{C}u^{D} \label{eq:evolution-of-B_AB}
\end{equation} 
\begin{equation}
\xi^{A}_{;B}u^{B} = B^{A}_{B} \xi^{B} \label{eq:evolution-deviation-vector}
\end{equation}
along with the geodesic equations (\ref{eq:geodesic-1} - \ref{eq:geodesic-4}). Here, $B_{AB} = \nabla_{A}u_{B}$ is gradient of velocity vector field, $\xi^{A}$ representing the deviation vector between two geodesics and Eq. (\ref{eq:evolution-deviation-vector}) is the evolution equation for the deviation vector. To see the evolution from the perspective of a local observer, we express the tensorial quantities in the frame basis. In coordinate and frame basis, the metric tensors and components of deviation vector are related via the vierbein field $e^{A}_{a}$ as
\begin{equation}
g^{AB} = e^{A}_{a}e^{B}_{b} \eta_{ab} ,\label{eq:relation-g-eta}
\end{equation}
\begin{equation}
\xi^{A} = e^{A}_{a} \xi^{a} .\label{eq:relation-deviation-A-a}
\end{equation}
Here capital and small latin indices stand for general spacetime coordinate (w.r.t coordinate basis) and local laboratory coordinate or local Lorentz frame (w.r.t. frame basis) respectively. 
As a first case, let us choose the boundary conditions (on $B_{AB}$) such that all the rotation components vanish (i.e. $\O_{AB}=0$). We choose a congruence of four geodesics (along with a reference geodesic at the {\em origin}) such that the cross-sectional area projected on $\sqrt{g_{11}} \xi^{1}$-$\sqrt{g_{33}} \xi^{3}$ or $\sqrt{g_{11}} \xi^{1}$-$\sqrt{g_{44}} \xi^{4}$ planes is of square shape at $\lambda = 0$ (or at the throat as $ l(\l=0) = 0$ for crossing geodesics). Then Eq. (\ref{eq:evolution-deviation-vector}) is solved for the four deviation vectors which represent the distances/deviations of the four geodesics (at the four corners of the square) from a central geodesic (co-moving observer) which is at the centre of the imaginary square (in the figures below). 
The following figures show the evolution of the projected cross sectional area for both 4D-GEB and 5D-WGEB geometries with respect to $\lambda$ (as geodesic congruence crosses the throat at $\lambda=0$). Various subcases correspond to different values of the wormhole parameter $m = 2, 4, 6$ and growing/decaying warp factor is analysed. In case where $m=4$ and $m=6$ geometries show similar behaviour we have presented evolution for $m=4$. Note that we choose the angular momentum `$h$' to be non-zero (and consistent with all the constraints), in this section, which plays an interesting role in 5D scenario.
 
\subsection{Case-1: 4D-GEB spacetime}

Congruences in the 4D-GEB spacetime is shown in Fig. (\ref{fig:4d-deviation-evolution}). 
The boundary values used in this subsection is shown in Appendix \ref{app-4}. 
Four snapshots of the area with increasing and decreasing $\l$ is depicted in left and right plots respectively. These plots show features of expansion and shear with no rotation as expected. The cross sectional area focuses as geodesic congruence approaches towards the throat $l=0$, while distortion in projected area decreases until it hits the throat and increases after passing through the throat. 
\begin{figure}[h]
\centering
\includegraphics[scale=.60]{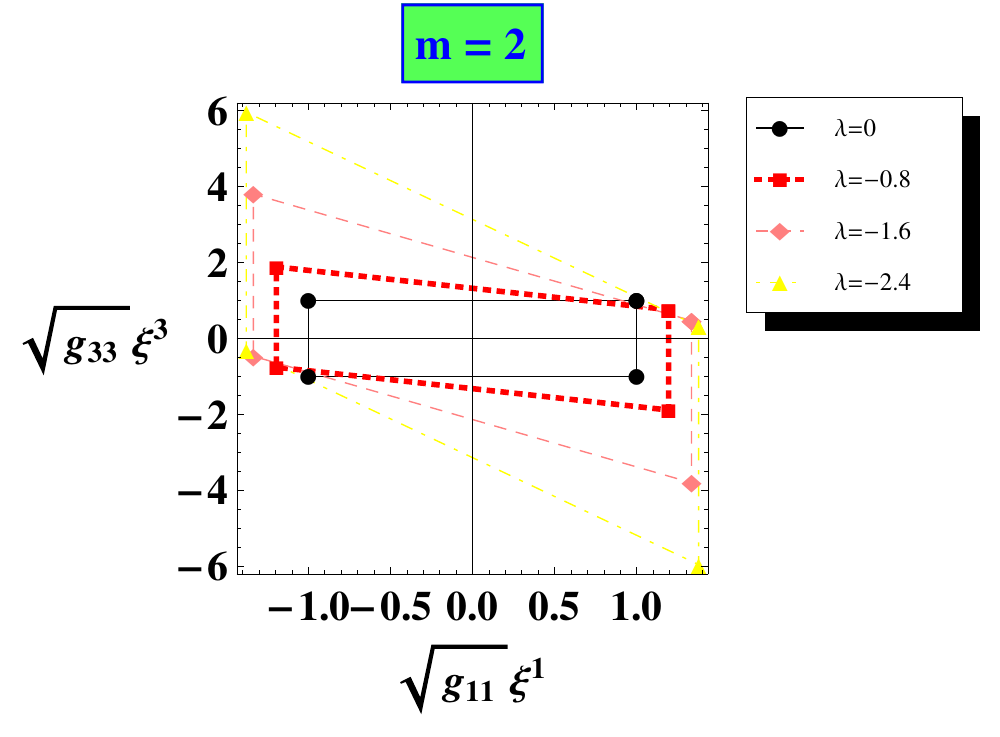}
\hspace{1cm}
\includegraphics[scale=.60]{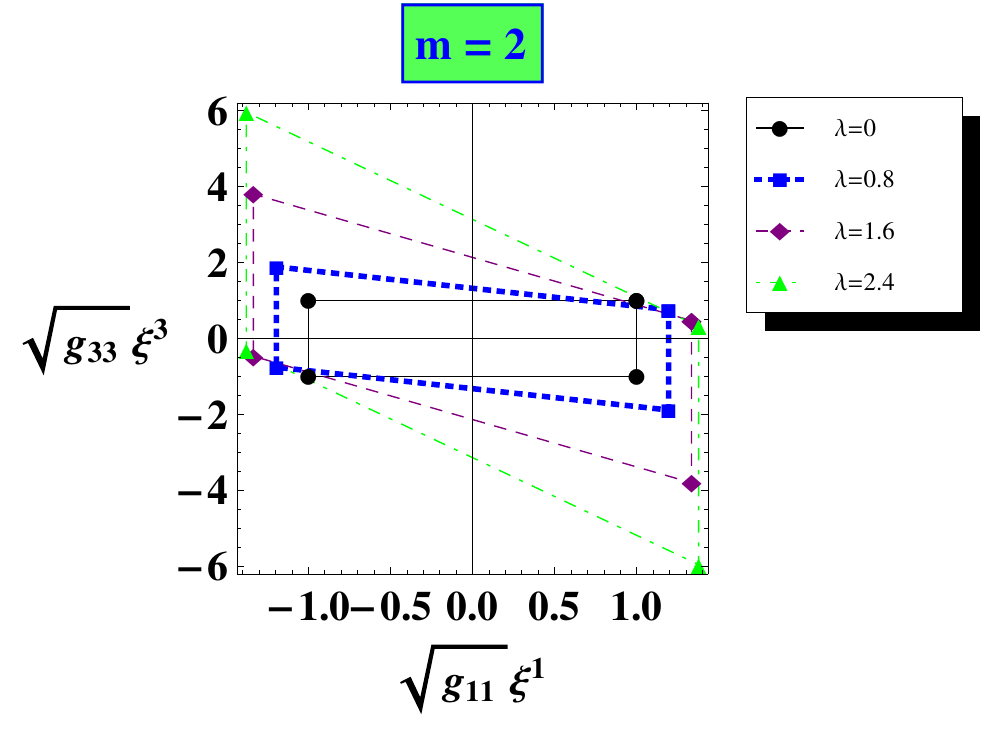}
\hspace{1cm}
\includegraphics[scale=.60]{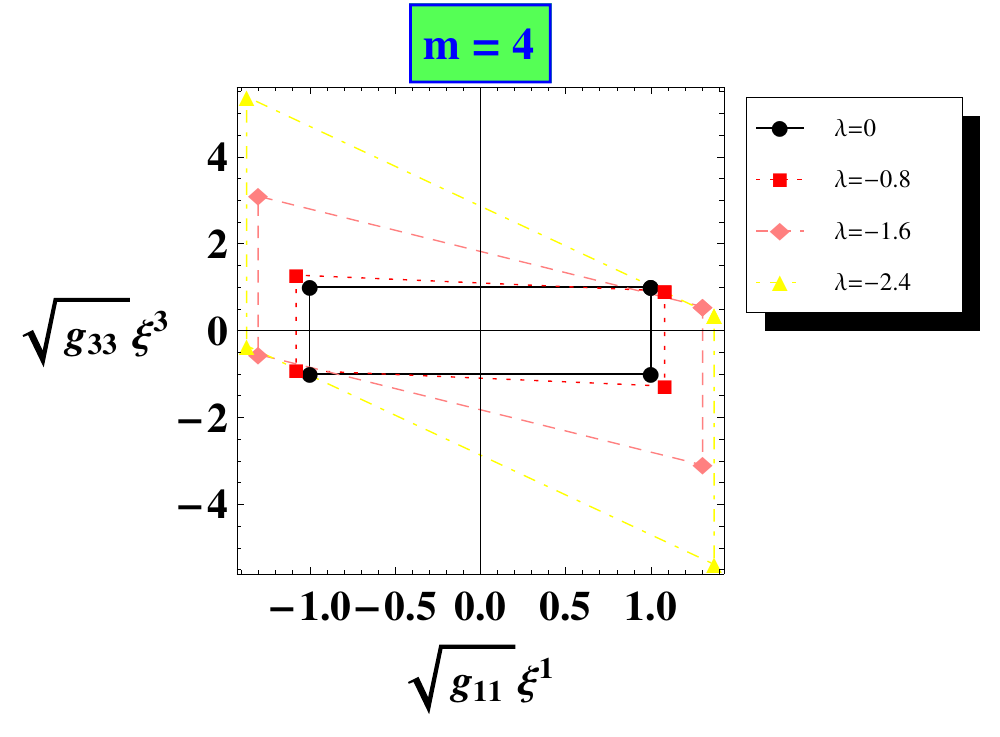}
\hspace{1cm}
\includegraphics[scale=.60]{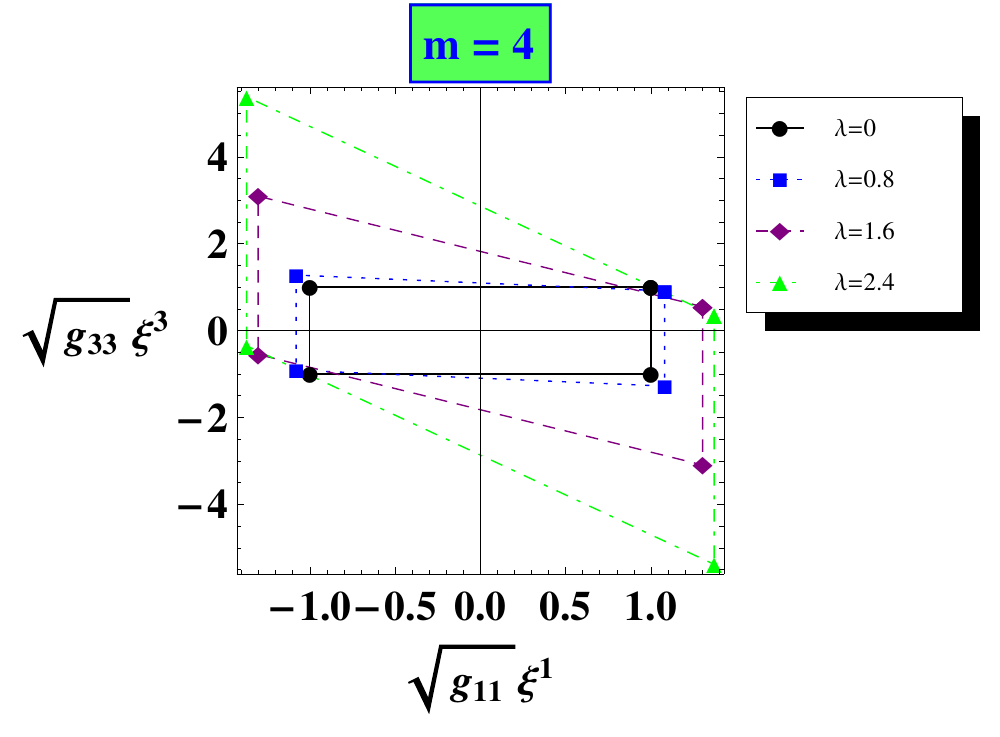}
\caption{Showing evolution of projected geodesic congruence plane (without rotation) in case of crossing congruences with different choice of wormhole parameter $m = 2, 4$ and $b_{0} = 1$. }
\label{fig:4d-deviation-evolution}
\end{figure} 

The evolution of cross sectional area is symmetric about the throat which is opposite to the overall expansion profile as shown in the the previous section.  
However, there are no qualitative difference in the evolution of cross-section between $m = 2$ and $m > 2$ geometries. Apparently, a congruence which is widespread and largely distorted coming from one asymptotic region crossing the throat with minimum size and distortion and regenerates its original shape and size on the other side. 

\subsection{Case-2: 5D-WGEB spacetime with growing warp factor}

\begin{figure}[H]
\centering
\includegraphics[scale=.6]{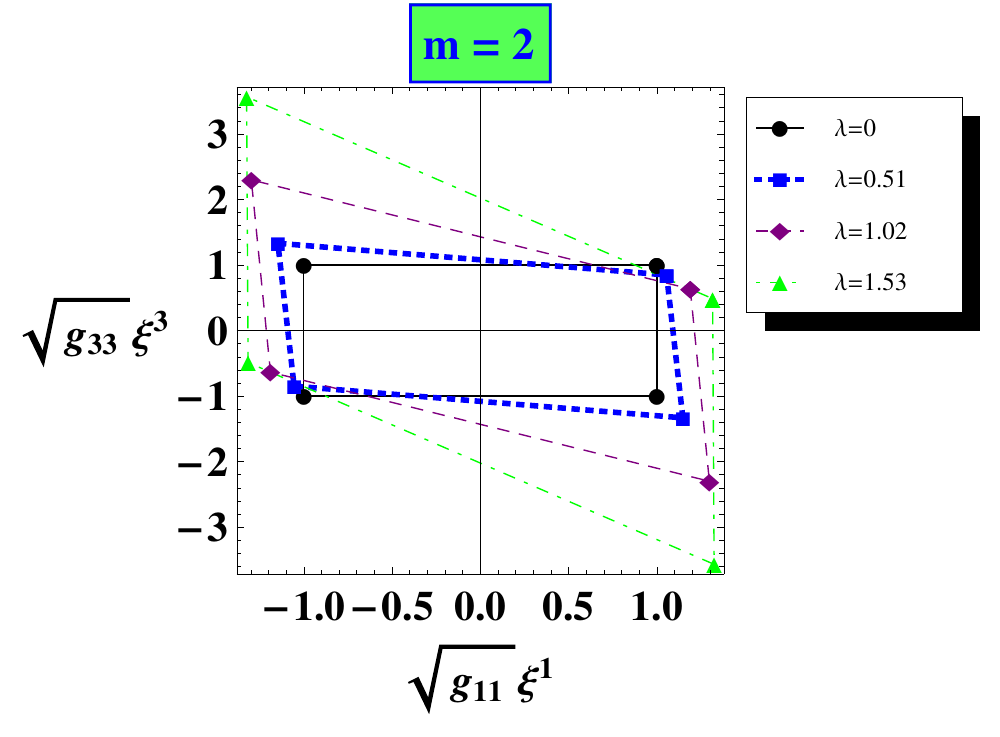}
\hspace{1cm}
\includegraphics[scale=.6]{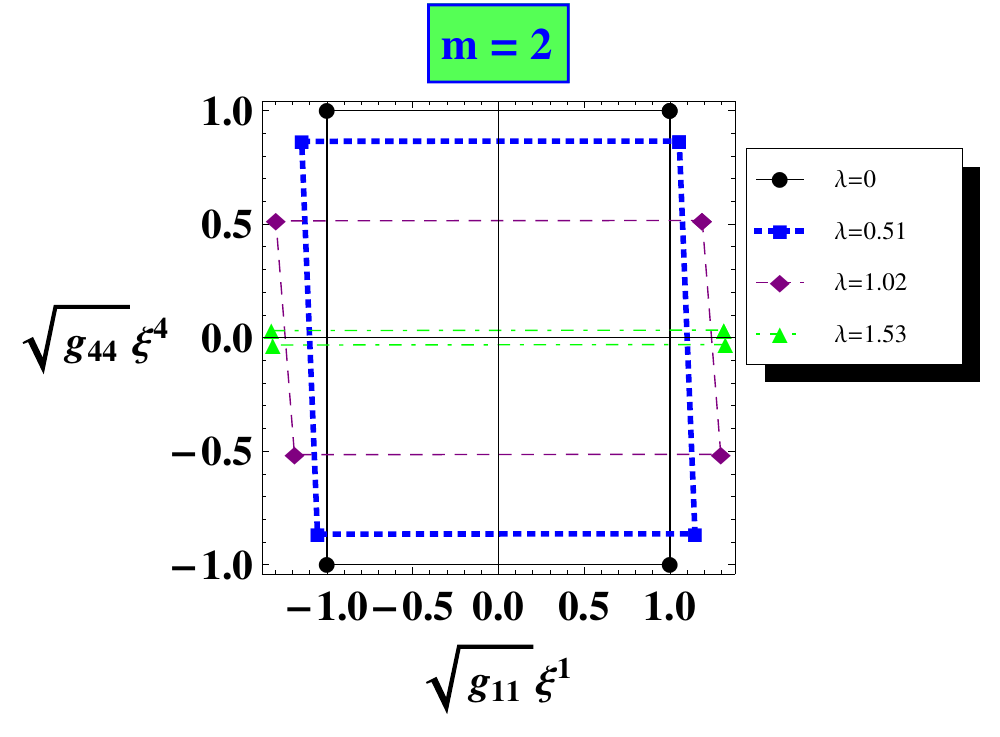}
%
%
\includegraphics[scale=.60]{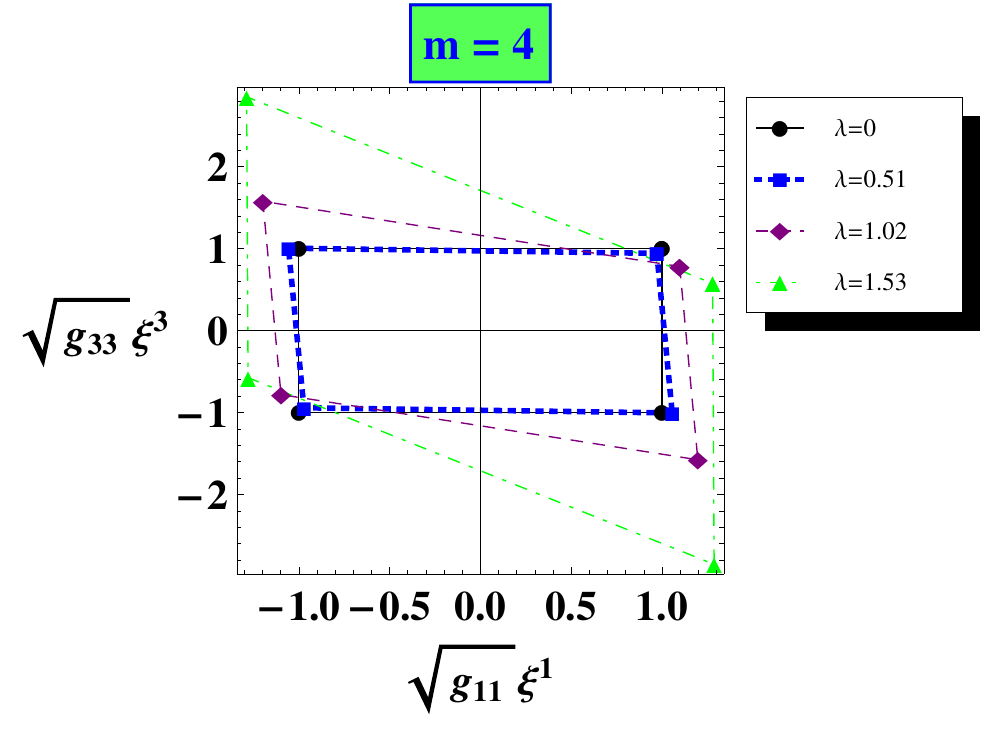}
\hspace{1cm}
\includegraphics[scale=.60]{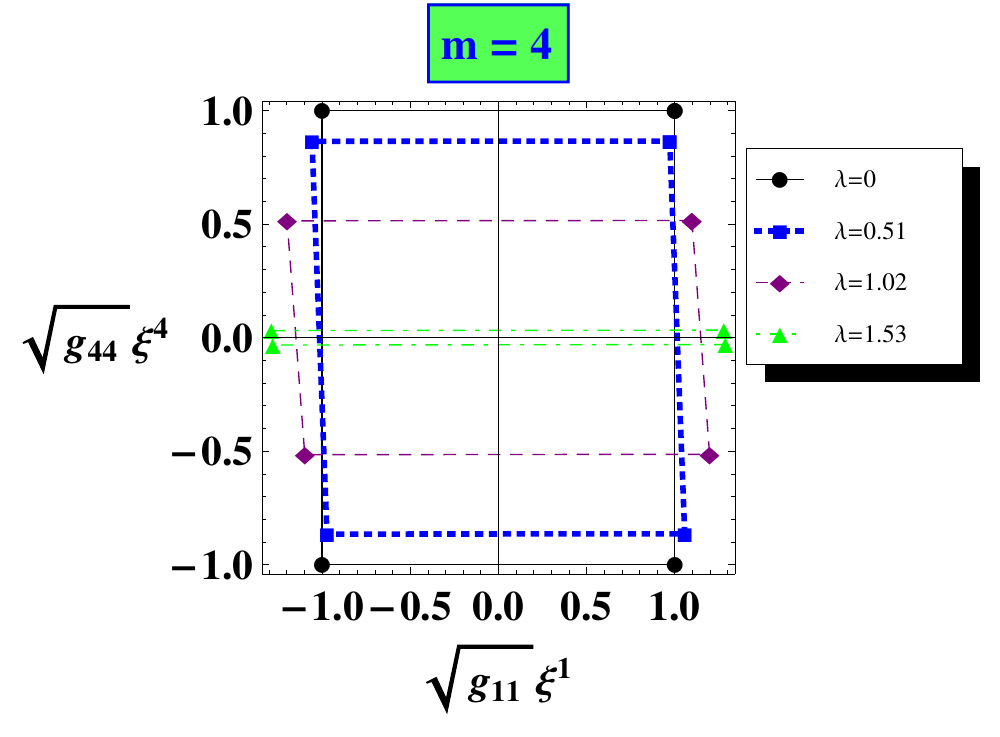}
\caption{Showing evolution of projected geodesic congruence plane (without rotation) in case of crossing congruences with growing warp factor for wormhole parameter $m =2, 4$ and $b_{0} = 1$. }
\label{fig:5d-grow-deviation-evolution}
\end{figure}

The evolution profiles of congruence cross-sections projected on the $\sqrt{g_{11}} \xi^{1}$-$\sqrt{g_{33}} \xi^{3}$ and $\sqrt{g_{11}} \xi^{1}$-$\sqrt{g_{44}} \xi^{4}$ plane for 5D-WGEB model with growing warp factor is shown in the Fig. (\ref{fig:5d-grow-deviation-evolution}) for two different values of $m = 2, 4$.
The boundary values used in this subsection is shown in Appendix \ref{app-5}. 
Since the evolution is symmetric about $\l=0$ (or $l=0$) we only present here plots for $\l \geq 0$ (or $l \geq 0$). 

The evolution of the area element projected on $\sqrt{g_{11}} \xi^{1}$-$\sqrt{g_{33}} \xi^{3}$ plane has almost similar profile as the 4D case with a crucial difference-- the area has rotation (that slowly increases with increasing $m$) too along with expansion and shear even though the rotation was zero at the throat as in the 4D case. Thus the warping factor triggers a rotation in the congruence if individual geodesics has angular momenta. In other words, congruence of geodesics with angular momenta that has initial rotation at asymptotic regions may become rotation-less while passing through the throat. 
The other effect of the growing warp factor is visible through the projection of the evolution on the $\sqrt{g_{11}} \xi^{1}$-$\sqrt{g_{44}} \xi^{4}$ plane which shows that the geodesics whose $y$-coordinate are some distant apart at $\l=0$, have the same $y$-coordinate in the asymptotic regions, i.e. separation along $y$-axis vanishes among the geodesics as $\l \ra \pm \infty$.  
However, there is no rotation of the congruence on this plane.

%
%

\subsection{Case-3: 5D-WGEB spacetime with decaying warp factor}

\begin{figure}[H]
\centering
\includegraphics[scale=.6]{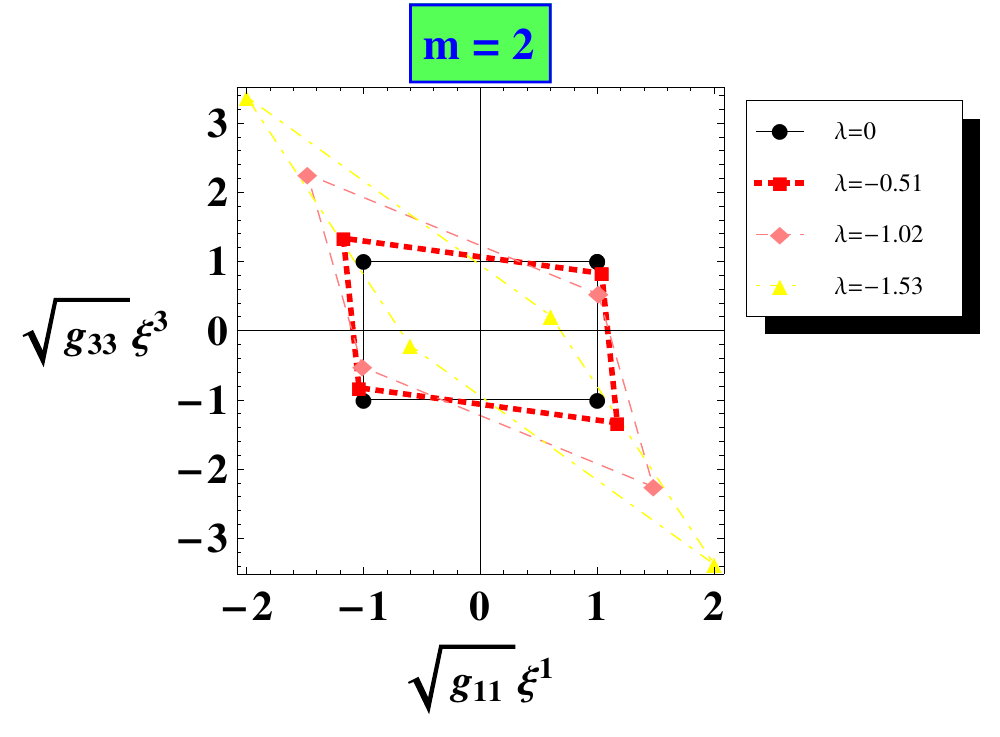}
\hspace{0.2cm}
\includegraphics[scale=.6]{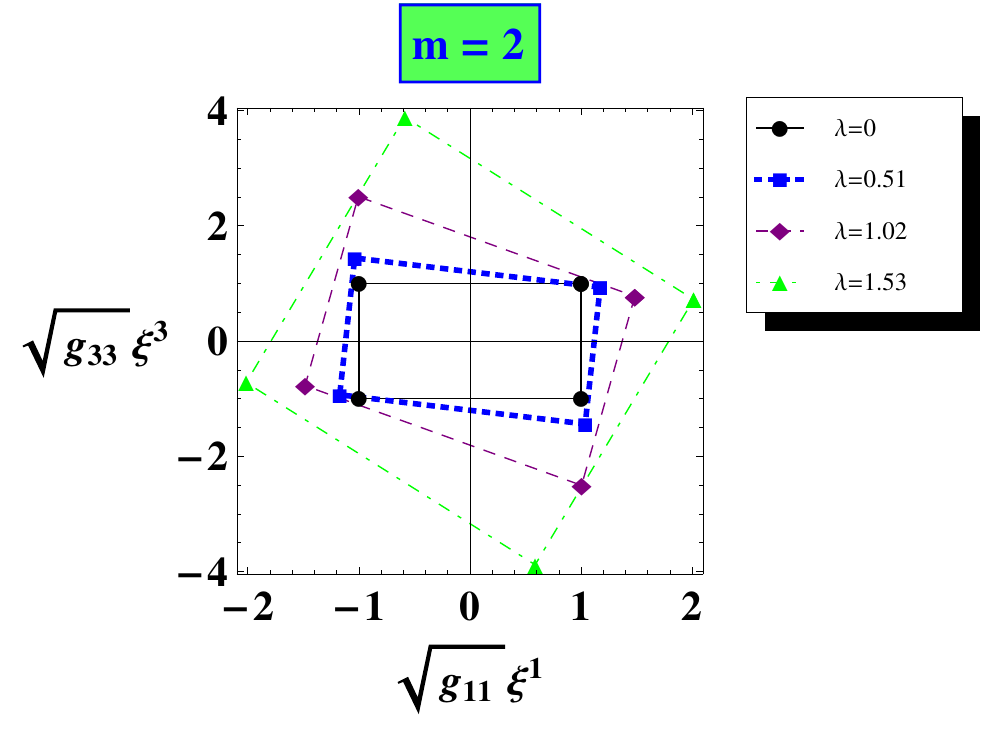}
\includegraphics[scale=.6]{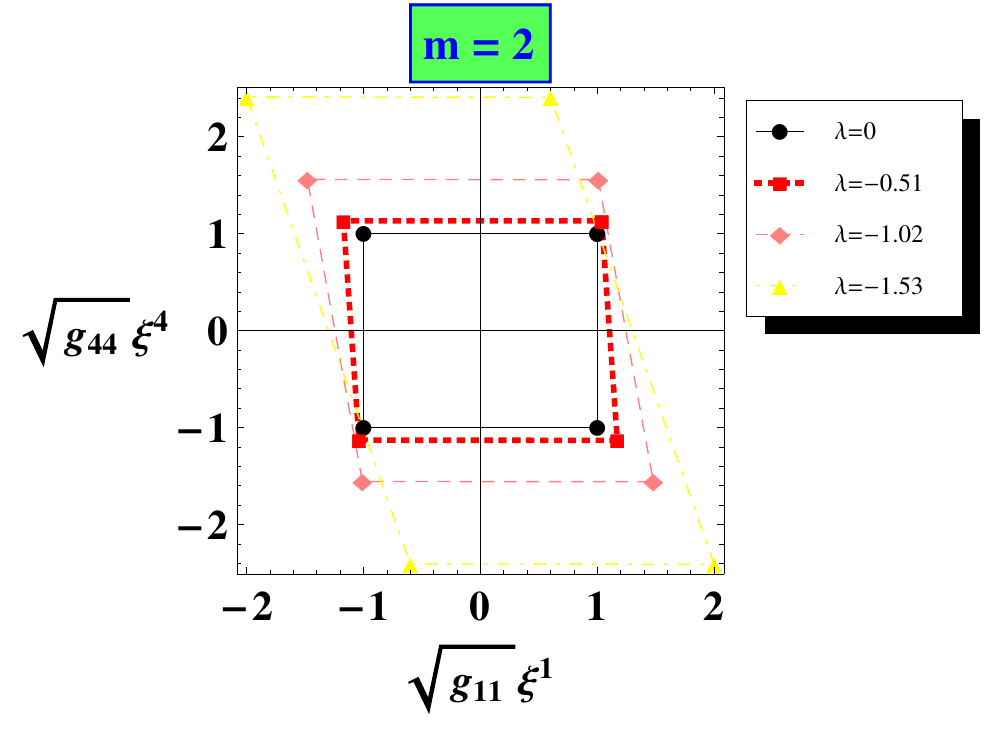}
\hspace{0.2cm}
\includegraphics[scale=.6]{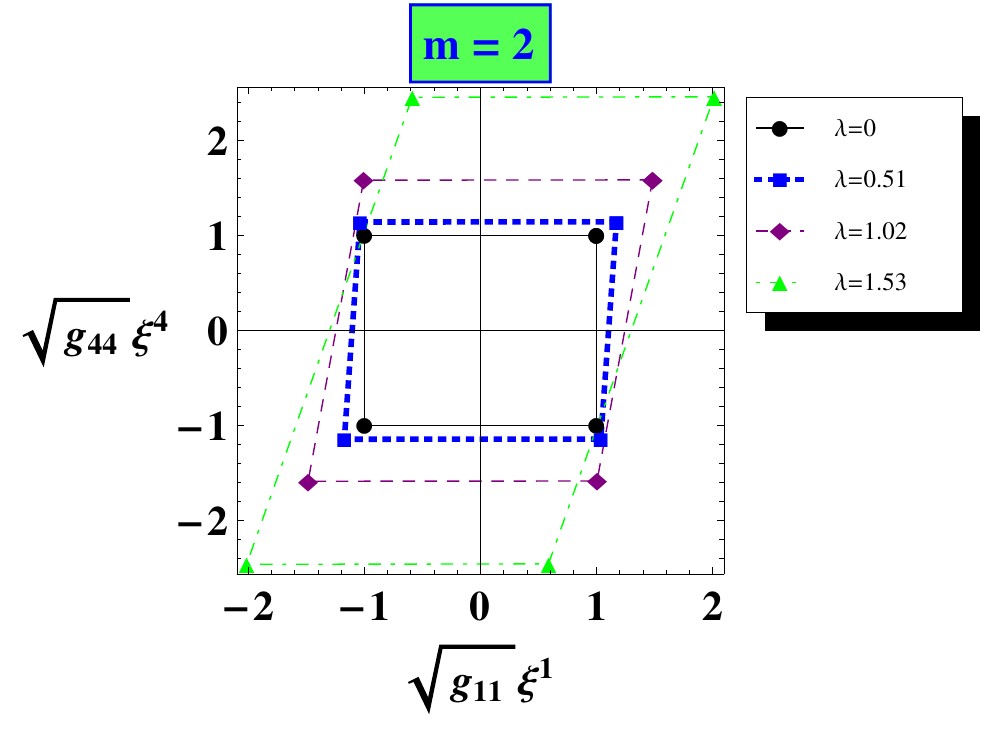}
\caption{Showing evolution of projected geodesic congruence plane (without rotation) in case of crossing congruences with decaying warp factor for wormhole parameter $m = 2$ and $b_{0} = 1$. }
\label{fig:5d-decay-deviation-evolution-2}
\end{figure}

\begin{figure}[H]
\centering
\includegraphics[scale=.60]{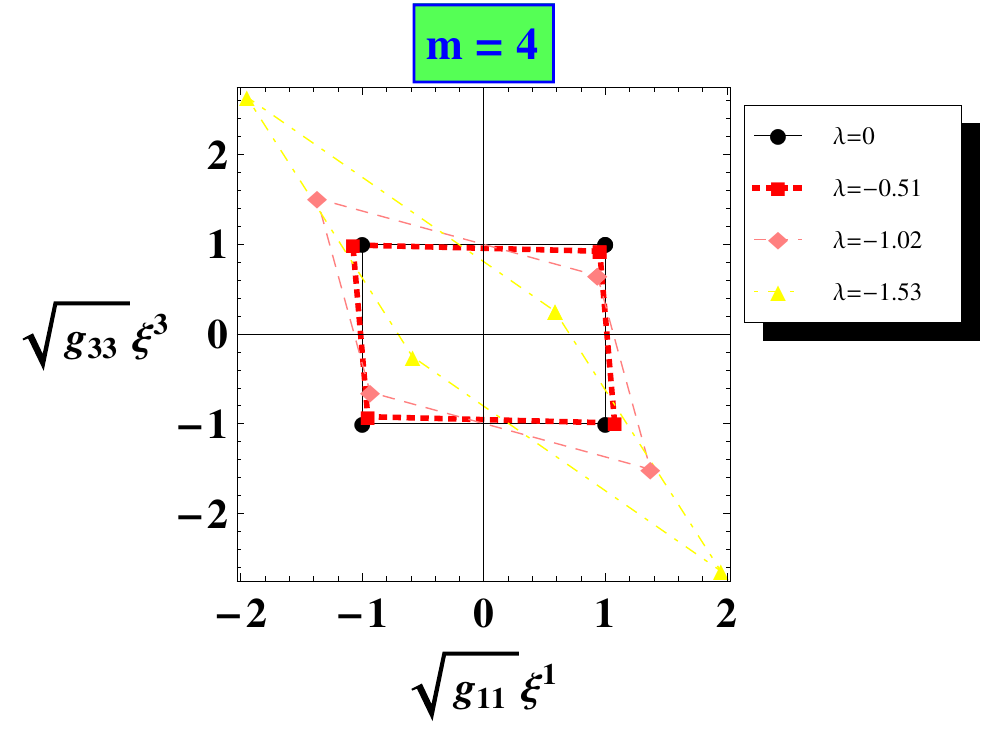}
\hspace{0.2cm}
\includegraphics[scale=.60]{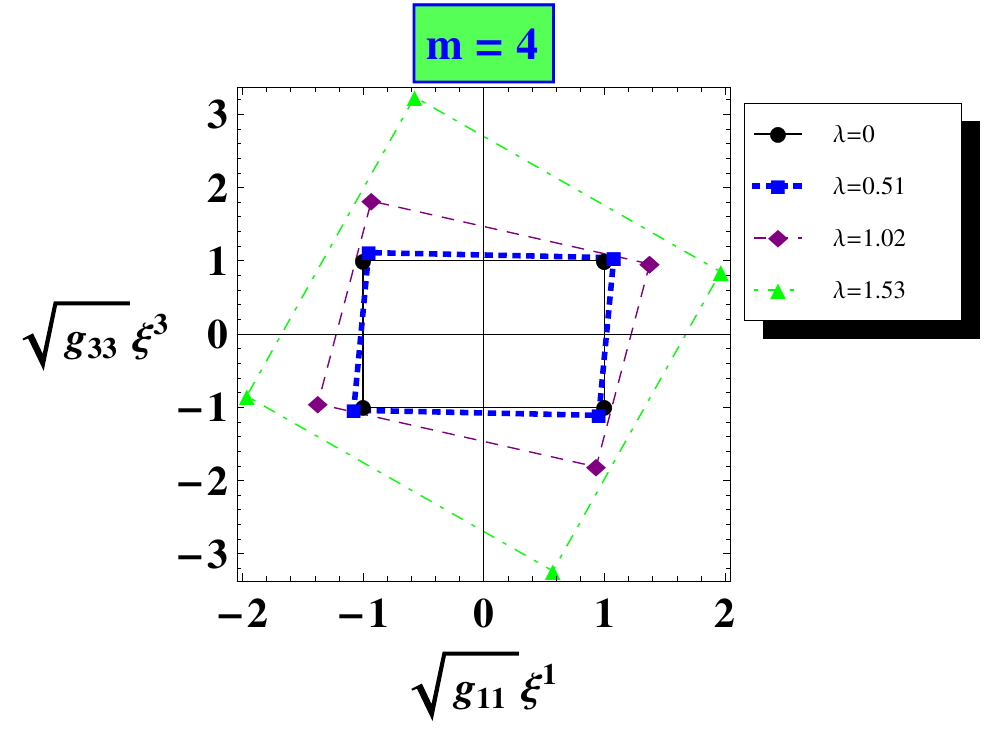}
\includegraphics[scale=.60]{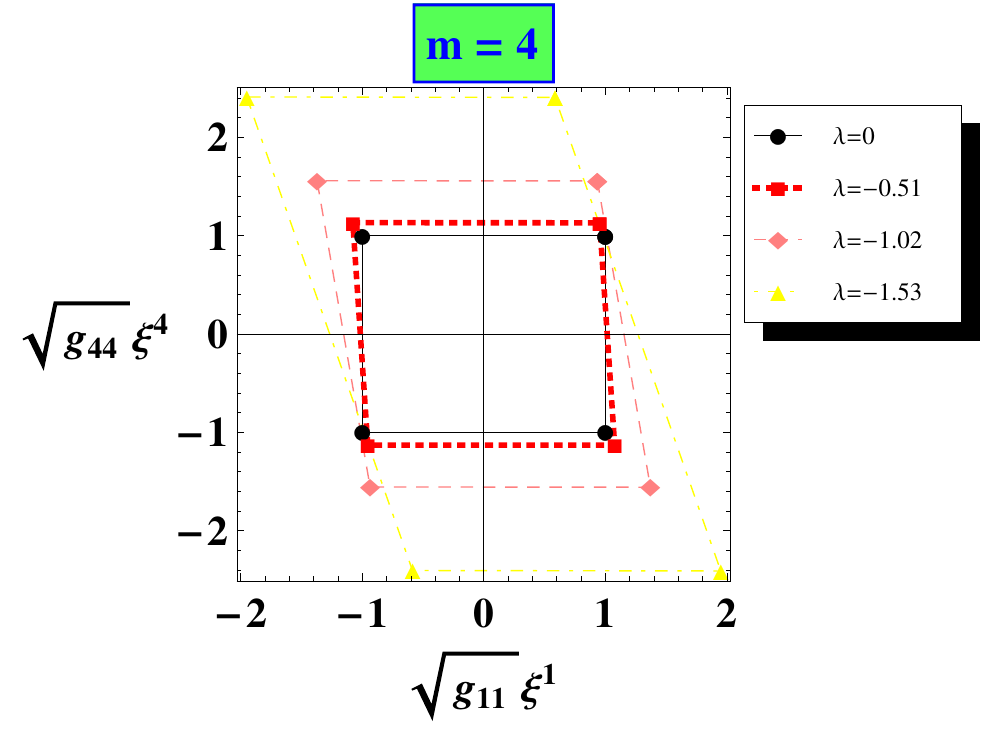}
\hspace{0.2cm}
\includegraphics[scale=.60]{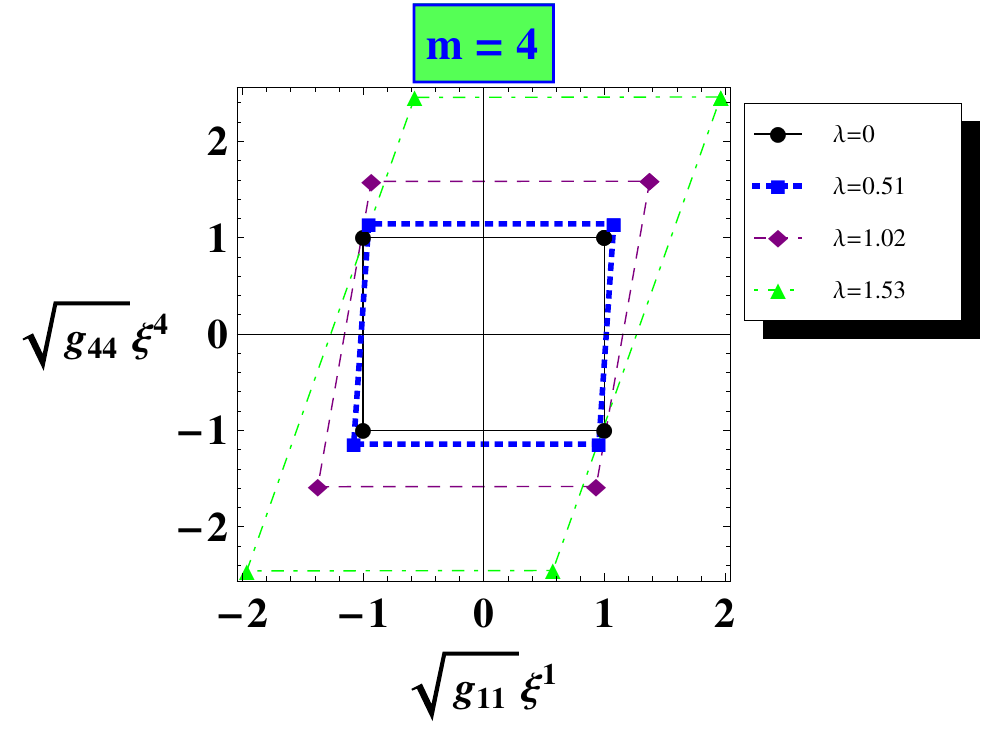}
\caption{Showing evolution of projected geodesic congruence plane (without rotation) in case of crossing congruences with decaying warp factor for wormhole parameter $m = 4$ and $b_{0} = 1$. }
\label{fig:5d-decay-deviation-evolution-4}
\end{figure}

The figures (\ref{fig:5d-decay-deviation-evolution-2}) and (\ref{fig:5d-decay-deviation-evolution-4}) show the evolution of projected area element with different $m$ values ($m = 2, 4$) in $\sqrt{g_{11}} \xi^{1}$-$\sqrt{g_{33}} \xi^{3}$ and $\sqrt{g_{11}} \xi^{1}$-$\sqrt{g_{44}} \xi^{4}$ plane for the 5D-WGEB model with a decaying warp factor. 
The boundary values used in this subsection is shown in Appendix \ref{app-6}. 
There are few factors remarkably different from the case with a growing warp factor.
First, evolution is not similar before and after crossing the throat. So we keep plots for both positive (on right) and negative (on left) $\l$.
Second, the presence of the decaying warp factor has amplified the magnitude of rotation of the area element considerably when $\l \ra \infty$ whereas rotation is almost negligible when $\l \ra -\infty$. This means a congruence with low rotation become highly rotating after crossing the throat and vice versa.

In contrast to the model with a growing warp factor, the evolution of the area element projected on the $\sqrt{g_{11}} \xi^{1}$-$\sqrt{g_{44}} \xi^{4}$ plane in the case of decaying warp factor has the opposite profile. 
Further, these figures reveal that the wormhole parameter, $m$, affects the area element in the same way for all the models, i.e. the magnitude of the expansion and distortion decreases as $m$ grows. 
Perhaps the most significant aspect of this section is the unique perspective that let us look into the details of the congruence rotation or {\em accretion}.




\section{Discussion} \label{sec:6}

Earlier we showed that the generalised 4D-Ellis-Bronnikov spacetime  embedded in warped 5D bulk satisfies the energy conditions thus is a viable wormhole model within the framework of general relativity. We further discussed the particle trajectories in 4D-GEB and 5D-GEB models and compared them. These investigations show that even the simple 5D-EB model ($m=2$ case) has nice properties as a wormhole. 
This work can be considered as the third article in this series where we have analysed congruence of timelike crossing geodesics in detail which is essential to understand the viability of travelling through these class of wormholes.
We derived the evolution of the so-called ESR variables for both the 4D-GEB and 5D-WGEB (with growing or decreasing warp factor) spacetimes by solving geodesic equations, geodesic deviation equations and the Raychaudhuri equation simultaneously and compared the results. The key findings of this analysis are as follows.

\begin{itemize}

\item For 4D-GEB spacetime, analytic expressions of ESR variables of the congruence were first determined without rotation ($\Omega_{AB} = 0$). The resulting plots reveal that $m = 2$ and $m > 2$ geometries can be easily distinguished based on the evolution of expansion scalar $\Theta$. ESR variables all vanish in a similar way in the asymptotic regions at $l \ra \pm \infty$. This essentially happens because the geometries are asymptotically flat. 
 
\item In the 4D-GEB geometry, for congruence with rotation, numerical results are consistent with the analytic calculation of ESR variables. Numerical analysis further show that the differences between ESR profiles in $m = 2$ and $m > 2$ geometries, that appear in absence of rotation, in fact disappear if rotation is present while the congruence is crossing the throat. 

\item The 5D-WGEB models are studied with growing and decaying warp factors. In absence of any rotation in the congruence, expansion and shear becomes very large in the asymptotic regions i.e. the congruence fails to retain its shape and size while crossing the throat. Remarkably, presence of rotation improves the situation for expansion scalar in case of growing warp factor and completely removes the divergences in case of decaying warp factor model. In fact the divergence in shear at asymptotic region is removed as well.
An universal feature, in absence of rotation, in both the 4D-GEB and 5D-WGEB models is that the expansion scalar corresponding to all values of $m$ has exactly same value at length scale $l = \pm b_{0}$.

\item In the snapshots of cross-sectional area of a congruence of timelike geodesics projected on 2D surfaces ($l-\phi$ and $l-y$ planes) provide another interesting perspective on the evolution of the ESR variables. This analysis clearly shows that, a congruence with zero rotation near the throat may have rotation in the asymptotic regions if the individual geodesic possesses angular momenta. This effect is considerably large in presence of decaying warp factor compared to the case of a growing warp factor. One may say, the warping factor essentially leads to {\em accretion}. Further an effect of increasing $m$ (which essentially means a steeper throat) has similar effect on the evolutions for all the 4D and 5D models.

\item It is important to note that when the warp factor is growing, geodesics have a turning point along the extra dimension \cite{Ghosh:2010gq}, which forces congruence singularity to occur, whereas when the warp factor is decaying, the occurrence of congruence singularity is completely dependent on boundary conditions. 
Our visualization analyses do show how the square elements change shape and get rotated because of variations in the metric functions and boundary conditions. We may contemplate such accretion effects for flows around brane-world wormholes (similar to braneworld black-hole scenario \cite{Pun:2008ua}). In such situations, it will become useful to pursue an approach very similar to what we have used in this article.

\end{itemize}

It still remains to analyse the congruences for {\em all} possible boundary conditions which should be imposed in the asymptotic regions. To avoid computational difficulty, in most cases, we imposed those conditions at the throat. Even then we could understand various effects the congruences will be subjected to while traversing through the hole. There are various other features yet to be studied for these class of wormholes regarding lensing, stability under perturbation as such. One crucial pointer is towards rotating Ellis-Bronnikov wormholes which seems to be possessing useful properties and yet not been addressed in the literature as much. We shall report on these issues in future communications.

\section{Appendix} 

\subsection{Boundary conditions to determine ESR profiles in 4D-GEB spacetimes} \label{app-1}

\noindent
$k = \sqrt{3}$, $h = 0 $; 
$t(0) = 0$, $l(0) = 0$, $\theta(0) = \pi/2$, $\phi(0) = 0$ ;\\
$\dot{t}(0) = k $, $\dot{l}(0) = 1.41421$, $\dot{\theta}(0) = 0$, $\dot{\phi}(0) = \frac{h}{(1 + l(0)^{m})^{2/m}} = 0 $ ;\\
$B_{AB}(0) = 0$ for without rotation and (initially $\theta$, $\Sigma_{AB}$ and $\Omega_{AB}$ is zero) ;\\
For with rotation case: $\theta$ and $\Sigma_{AB}$ are zero at $\l=0$ but non-zero $\Omega_{AB}$ are chosen such that $\Omega_{AB}u^{B} = 0 = u^{A}\Omega_{AB}$ at $\lambda = 0$ 

\subsection{Boundary conditions to determine ESR profiles in 5D-WGEB  spacetimes with growing warp factor}\label{app-2}

\noindent
$T = \sqrt{3}$, $H = 0 $; 
$t(0) = 0$, $l(0) = 0$, $\theta(0) = \pi/2$, $\phi(0) = 0$, $y(0) = 0.1 $ ;\\
$\dot{t}(0) = \frac{T}{e^{2f(y)}} = 1.71485 $, $\dot{l}(0) = 1.39665 $, $\dot{\theta}(0) = 0$, $\dot{\phi}(0) = \frac{H}{(1 + l(0)^{m})^{2/m}} = 0 $,$ \dot{y}(0) = 0 $ ;\\
$B_{AB}(0) = 0$ for without rotation that implies $\theta(0)$, $\Sigma_{AB}(0)$ and $\Omega_{AB}(0)$ is zero ;\\
For with rotation case: $\theta$ and $\Sigma_{AB}$ are zero at $\l=0$ but non-zero $\Omega_{AB}$ are chosen such that $\Omega_{AB}u^{B} = 0 = u^{A}\Omega_{AB}$ at $\lambda = 0$ 

\subsection{Boundary conditions to determine ESR profiles in 5D-WGEB  spacetimes with decaying warp factor}\label{app-3}
\noindent
$T = \sqrt{3}$, $H = 0 $ ;
$t(0) = 0$, $l(0) = 0$, $\theta(0) = \pi/2$, $\phi(0) = 0$, $y(0) = 0.1 $ ;\\
$\dot{t}(0) = \frac{T}{e^{2f(y)}} = 1.74943 $, $\dot{l}(0) = 1.43195 $, $\dot{\theta}(0) = 0$, $\dot{\phi}(0) = \frac{H}{(1 + l(0)^{m})^{2/m}} = 0 $,$ \dot{y}(0) = 0 $ ;\\
$B_{AB}(0) = 0$ for without rotation that implies $\theta(0)$, $\Sigma_{AB}(0)$ and $\Omega_{AB}(0)$ is zero ;\\
For with rotation case: initially $\theta$ and $\Sigma_{AB}$ are still zero but non-zero $\Omega_{AB}$ are chosen such that $\Omega_{AB}u^{B} = 0 = u^{A}\Omega_{AB}$ at $\lambda = 0$ 

\subsection{Boundary conditions for evolution of cross-sectional area in 4D-GEB spacetimes}\label{app-4}
\noindent
$k = \sqrt{3}$, $h = \sqrt{\frac{k^{2} - 1 }{2}} = 1$ ;
$t(0) = 0$, $l(0) = 0 $, $\theta(0) = \pi/2$, $\phi(0) = 0$ ;\\
$\dot{t}(0) = k$, $ \dot{l}(0) = 1$, $\dot{\theta}(0) = 0$, $\dot{\phi}(0) = \frac{h}{(1 + l(0)^{m})^{2/m}} = h = 1 $ ;\\
$\dot{l}(0)$ is calculated from the timelike geodesic constraint ;\\
$B_{AB}(0) = 0$ ;
$\xi^{1}(0) = \pm 1$, $\xi^{2}(0) = 0$, $\xi^{3}(0) = \pm 1$ ;\\
$\xi^{0}(0) = \pm \frac{2}{\sqrt{3}}, 0$ is calculated from $u_{A}\xi^{A} = 0$. 

\subsection{Boundary conditions to determine evolution of cross-sectional area in 5D-WGEB spacetimes with growing warp factor}\label{app-5}

\noindent
$t(0) = 0$, $l(0) = 0$, $\theta(0) = \pi/2$, $\phi(0) = 0$, $y(0) = a = 0.1$ ;\\
$\dot{t}(0) = T$, $ \dot{l}(0) = 0.98758 $ , $\dot{\theta}(0) = 0$, $\dot{\phi}(0) = \frac{H}{\cosh(a)^{2}(1 + l(0)^{m})^{2/m}} = 0.98758$, $\dot{y}(0) = b = 0$ ;\\
$T = \sqrt{3}$, $H = \sqrt{\frac{k^{2} - \cosh(a)^{2}(b^{2} + 1)}{2}} = 0.997489$ ;\\
$B_{\mu\nu}(0) = 0$ , where $\mu$ and $\nu$ runs from $0$ to $4$ ;\\
$\xi^{1}(0) = \pm \frac{1}{\cosh(a)} = \pm 0.995021$, $\xi^{2}(0) = 0$, $\xi^{3}(0) = \pm \frac{1}{\cosh(a)} = \pm 0.995021$, $\xi^{4}(0) = \pm \frac{1}{\cosh(a)} = \pm 0.995021$ ;\\
$\xi^{0}(0) = \pm 1.19213, \pm 0.0574475$ is calculated from $u_{A}\xi^{A} = 0$. 

\subsection{Boundary conditions to determine evolution of cross-sectional area in 5D-WGEB spacetimes with decaying warp factor}\label{app-6}

$t(0) = 0$, $l(0) = 0$, $\theta(0) = \pi/2$, $\phi(0) = 0$, $y(0) = a = 0.1$ ;\\
$\dot{t}(0) = T$, $\dot{l}(0) = 1.01254 $ , $\dot{\theta}(0) = 0$, $\dot{\phi}(0) = \frac{H}{sech(a)^{2} (1 + l(0)^{m})^{2/m}} = 1.01254$, $\dot{y}(0) = b = 0$ ;\\
$T = \sqrt{3}$, $H = \sqrt{\frac{k^{2} - sech(a)^{2}(b^{2} + 1)}{2}} = 1.00248$ ;\\
$\xi^{1}(0) = \pm \frac{1}{sech(a)} = \pm 1.005$, $\xi^{2}(0) = 0$, $\xi^{3}(0) = \pm \frac{1}{sech(a)} = \pm 1.005 = \xi^{4}(0)$ ; \\
$\xi^{0}(0) = \pm 1.23305, \pm 0.0580239 $ is calculated from $u_{A}\xi^{A} = 0$. 

\noindent{\bf Data Availability Statement:} Data sharing not applicable to this article as no datasets were generated or analysed during the current study.



\end{document}